\documentclass[fleqn,usenatbib]{mnras}

\usepackage{newtxtext,newtxmath}

\usepackage[T1]{fontenc}
\DeclareRobustCommand{\VAN}[3]{#2}
\let\VANthebibliography\thebibliography
\def\thebibliography{\DeclareRobustCommand{\VAN}[3]{##3}\VANthebibliography}
\def\simgt{\lower.5ex\hbox{$\; \buildrel > \over \sim \;$}}%
\def\simlt{\lower.5ex\hbox{$\; \buildrel < \over \sim \;$}}%
\usepackage[flushleft]{threeparttable}
\usepackage{deluxetable}
\usepackage{graphicx}	
\usepackage{amsmath}	
\usepackage{multicol}        
\usepackage{bm}		
\usepackage{pdflscape}

\newcommand\bvo{\mbox{$(B\,-\,V)_{0}$}}
\newcommand\vio{\mbox{$(V\,-\,I)_{0}$}}
\newcommand{\hi}{H\textsubscript{I}}%
\newcommand\meanbvo{\mbox{$\langle (B\,-\,V)_{0} \rangle$}}
\def\msun{M\textsubscript{\(\odot\)}}%
\newcommand\bv{\mbox{$B\,-\,V$}}
\def\vi{\mbox{$V\,-\,I$}}
\newcommand\notsotiny{\@setfontsize\notsotiny{6}{7}}
\newcommand{\tjf}{\normalfont}

\title[Dwarf Galaxy Discoveries from the KMTNet Supernova Program III.]
{Dwarf Galaxy Discoveries from the KMTNet Supernova Program III.
the Milky-Way Analog NGC~2997 Group}

\author[T. J. Fan et al.]
{
Tony Junjing Fan,$^{1}$
Dae-Sik Moon,$^{1}$
Hong Soo Park,$^{2,3,4}$
Dennis Zaritsky,$^{4}$
Sang Chul Kim,$^{2,3}$
\newauthor
Youngdae Lee,$^{5}$
Ting S. Li,$^{1}$
Yuan Qi Ni,$^{1}$
Jeehye Shin,$^{2,3}$
Sang-Mok Cha,$^{2,6}$
Yongseok Lee,$^{2,6}$
\\
$^{1}$David A. Dunlap Department of Astronomy and Astrophysics, University of Toronto, 50 St. George Street, Toronto, ON M5S 3H4, Canada\\
$^{2}$Korea Astronomy and Space Science Institute, 776, Daedeokdae-ro, Yuseong-gu, Daejeon 34055, Republic of Korea\\
$^{3}$Korea University of Science and Technology (UST), Daejeon 34113, Republic of Korea\\
$^{4}$Steward Observatory, University of Arizona, 933 North Cherry Avenue, Tucson, AZ 85719, USA\\
$^{5}$Department of Astronomy and Space Science, Chungnam National University, Daejeon 34134, Republic of Korea\\
$^{6}$School of Space Research, Kyung Hee University, Yongin 17104, Republic of Korea
}

\date{Accepted XXX. Received YYY; in original form ZZZ}

\pubyear{2022}

\begin{document}

\label{firstpage}
\pagerange{\pageref{firstpage}--\pageref{lastpage}}
\maketitle

\begin{abstract}
We present the discovery of 48 new and the analysis of 55, 
including 7 previously discovered, dwarf galaxy candidates 
around the giant spiral galaxy NGC~2997 using deep $BVI$ images 
from the KMTNet Supernova Program. 
Their $V$-band central surface brightness and total absolute magnitudes are
in the range of 20.3--26.7 mag arcsec$^{-2}$ 
and --(8.02--17.69) mag, respectively, 
while the $I$-band effective radii are 0.14--2.97 kpc.
We obtain $\alpha$ $\simeq$ --1.43 $\pm$ 0.02
for the faint-end slope of their luminosity function,
comparable to previously measured values but
shallower than theoretical predictions 
based on $\Lambda$CDM models.
The distance-independent distributions of their mass and color 
suggest {\tjf that the group could have recently accreted new massive members  from the surrounding fields.}
The systematically bluer colors of the brighter members 
indicate younger stellar population and higher star formation 
activities in them, which appears to be consistent with similar findings 
from the SAGA or ELVES survey. 
We suggest that the massive and bluer dwarf galaxies in the group
{\tjf have experienced less environmental quenching due to their recent accretion,}
while environmental quenching is more effective for the low-mass members.
The interpretation of NGC~2997 being populationally young with recent accretion of massive members
is also consistent with the overall morphological distribution 
of the dwarf galaxies showing a lack of morphologically evolved candidates 
but a plethora of irregularly shaped ones.
Our detection rate of dwarf galaxy candidates in the NGC~2997 group
and their inferred star formation activities are 
comparable to those found in Milky Way analog systems 
from recent surveys within the magnitude limit 
M$_{V}$ $\lesssim$ --13 mag.
\end{abstract}




\section{Introduction}\label{sec:intro}

The study of dwarf galaxies plays a crucial role in our 
understanding of several important issues in astronomy.
It provides critical insights into the processes through which 
galaxy formation/evolution occurs, as well as observational tests 
for the distribution of dark matter and the $\Lambda$CDM model.
For example, studies of dwarf galaxies based on large surveys 
have shown that satellite galaxies form in more than one channel 
\citep[e.g.][]{Lisker2007}, and dwarf galaxies appear to show 
an extreme range in the amount of dark matter content,
from being almost free of dark matter to being almost entirely made
of dark matter \citep[e.g.][]{2016vanDokkum,2021Khalifeh}.
Despite recent efforts, there exists a substantial gap 
in our understanding of the origin and properties of dwarf galaxies, 
such as the well-known discrepancies in central densities and abundance
of dwarf galaxies between observational data on the Milky Way system 
and the $\Lambda$CDM model predictions,
known as the ``small-scale challenges'' \citep{2015Weinberg, 2021Mao},
or whether ``galactic conformity" is universal
in which both central host galaxies and satellite galaxies
show similar properties in star-formation or colors 
\citep{2006Weinmann,2020Otter,2021Poulain}.

The availability of recent wide-field deep survey 
data has made it possible to study 
dwarf galaxies and their properties
beyond the Milky Way (MW) and the Local Group 
\citep[$\simlt$ 3 Mpc;][]{2018Redd},
furthering our understanding of their luminosity functions (LFs)
and star formation activities in nearby groups.
Dwarf galaxies identified in the fields of
M81 \citep[at 3.6 Mpc;][]{2009Chiboucas}, 
NGC~5128 \citep[at 3.8 Mpc;][]{2016Crnojevic},
M83 \citep[at 4.0 Mpc;][]{2015Muller}, 
M101 \citep[at 7.0 Mpc;][]{2019Bennet}, 
NGC~2784 \citep[at 9.8 Mpc;][]{2017Park}, 
M96 \citep[at 10.7 Mpc;][]{2018Muller},
and NGC~3585 \citep[at 20.4 Mpc;][]{2019Park} 
have extended LFs of dwarf satellites 
down to M$_{V}$ $\simeq$ --7 mag, 
showing discrepancies that the observed faint-end slopes are still 
consistently flatter than the $\Lambda$CDM model predictions 
\citep[e.g.,][]{1999Klypin, 2002Trentham, 2008Springel, 2016Han, 2019PengfeiLi}.
The recent SAGA survey \citep{2017Geha,2021Mao}, 
{\tjf ELVES survey \citep{2021ELVES,2022ELVES},}
as well as studies of nearby groups such as 
M101 \citep{2019Bennet} and M94 \citep{2020Bennet}, on the other hand,
have shown that star formation in the faint (M$_{V}$ $\simgt$ --12 mag) 
satellite galaxies of MW-analog systems at $\simlt$ 41 Mpc 
is significantly less active than the massive ones,
providing opportunities to understand the effects of 
mass and environment on quenching processes \citep[e.g.,][]{2022Samuel}.
In addition, some studies have revealed the presence 
of apparent magnitude ``gaps'' or ``humps'' in some of the observed LFs:
the former wherein no galaxies found \citep[e.g.,][]{2019Bennet}, while 
the latter with sudden increase of the LF slopes
\citep[e.g.,][]{2019Park}.
They also show a morphological diversity in dwarf galaxies,
ranging from ultra-diffuse galaxies (UDGs) to nucleated dwarf galaxies
\citep[e.g.,][]{2015vanDokkum,2017Park,2018Muller,2019Bennet,2019Park}.
The origin of these features may bear important clues 
to our understanding of gravitational interactions 
among the members and the host of a galaxy group 
as well as the environmental effects to their evolutionary paths.

We have recently begun a systematic search of dwarf galaxies
around more than 100 nearby galaxies observable 
in the Southern Hemisphere using deep stacked images from 
the Korea Microlensing Telescope Network
\citep[KMTNet,][]{2016Kim}
Supernova Program \citep[KSP,][]{2016Moon}.
By stacking several hundred, sometimes exceeding 1000, 
60-s $BVI$ images obtained with the three wide-field 
(= 2\degr $\times$ 2\degr\ at a pixel sampling of 0\farcs4\ per pixel)
1.6-meter telescopes located in Chile, South Africa and Australia
that are primarily dedicated to detecting supernovae and optical transients
during their early phases
\citep[e.g.,][]{2019Afsariardchi,2021Moon,2022Lee,2022Ni,2023arXivNi,2023Ni},
we expect to discover a substantial number of new dwarf galaxies 
on images that readily reach a limiting surface 
brightness of $\simgt$ 28 mag arcsec$^{-2}$.
In our previous two studies of the NGC~2784 
and NGC~3585 groups using the KSP data,
we reported discoveries of 30 (for NGC~2784) and 46 (NGC~3585)
new dwarf galaxies \citep{2017Park,2019Park}.
One particular advantage of the KSP-data based studies of dwarf galaxies 
is the homogeneity of data obtained with three colors of $BVI$, 
which can enable unprecedented systematic comparative studies 
and statistical analyses of the discovered dwarf galaxies 
in different groups/clusters.
This can potentially lead to in-depth understanding of 
how different types of host galaxies and environments
influence differently to the evolution of satellite galaxies 
in their structural, morphological and dynamical features. 

In this paper, we present our KSP based discoveries 
of dwarf galaxies in the field of NGC~2997,
which are our first dwarf galaxy discoveries around 
a spiral-type host galaxy using the KSP data.
NGC~2997 is a giant spiral galaxy of type SAB(rs)c 
with a redshift of z $\simeq$ 0.003611 \citep{2004Meyer}
and an inclination angle of 40\degr \citep{1981Milliard}.
It is the largest member of the loose galaxy group LGG~180 
\citep{1993Garcia} and is relatively isolated from other major galaxies
with no known companion within $\sim$ 110 kpc \citep{2009Hess}.
The galaxy is reported to have a dynamical mass of 
2.1 $\times$ 10$^{11}$ \msun, 
a gas accretion rate of 1.2 \msun\ yr$^{-1}$,
and a thick H~I disk \citep{2009Hess}.
It has an $I$-band apparent magnitude of $I$ $\simeq$ 9.88 mag
\citep{2005Doyle}, and a $K$-band apparent magnitude of $K$ $\simeq$ 6.41 mag
\citep{2006Skrutskie}. 
The distance to NGC 2997 has been estimated to be 
in the range of 9.5--14.8 Mpc, including:
(1) 9.5 Mpc using tertiary distance indicators 
(magnitudes, luminosity index, and diameters)
\citep{1979deVaucouleurs,1999Larsen&Richter},
(2) 14.8 Mpc from the multi-attractor velocity flow model
\citep{2005Masters}
with $H_{0}$ = 72 km s$^{-1}$ Mpc$^{-1}$ \citep{2011Pisano}, 
and (3) 12.2 $\pm$ 0.9 Mpc based on 
the near-infrared Tully-Fisher relation \citep{2009Hess}.
We adopt the distance of 12.2 Mpc from \citet{2009Hess} in this work.
Its absolute $K$-band magnitude is --24.02 mag, which is within 
the range of --23 to --24.4 mag of the MW-analog galaxies
studied in the SAGA \citep{2017Geha,2021Mao} 
or ELVES survey \citep{2021ELVES, 2022ELVES}. 
Hence, this provides an opportunity of comparing dwarf galaxies 
identified in another MW-analog host at a closer distance 
with much deeper limiting magnitudes 
than those in the SAGA survey at $>$ 25 Mpc.
In \S\ref{sec:obs}, we provide the details of our observations and 
data reduction process, while we describe the identification of 
dwarf galaxy candidates (DGCs) and how we measure their 
photometric and morphological properties in \S\ref{sec:DGC}. 
\S\ref{sec:group} details group properties of the DGCs,
including their radial distributions, color and structural parameters, 
and luminosity function.
We discuss in \S\ref{sec:DE} that the observed features 
of the NGC~2997 group are compatible with {\tjf a scenario that
the group has recently been accreting massive, blue dwarf galaxies
from the surrounding fields.}
We summarize our conclusions in \S\ref{sec:sum}.

\section{Observations and Data Reduction}\label{sec:obs}

As part of the KSP (see above), 
we had monitored two 4 $\square\degr$ fields around the 
giant spiral galaxy NGC~2997 between February 2017 and June 2019.
The wide field (= 2\degr\ $\times$ 2\degr) CCD cameras of 
the KMTNet consist of four E2V CCD chips, 
with each chip covering 1 $\square\degr$ field
at a sampling size of 0\farcs4 per pixel.
There is a gap of 3\farcm1 between the chips in the east-west direction, 
while that in the north-south direction is 6\farcm2. 
Figure~\ref{fig:MAP} shows the distribution of the two observed fields 
in $I$-band, composed of eight sub-fields of 1 $\square\degr$,
with two sub-fields largely overlapping 
near the center of the entire field,
where the host galaxy NGC~2997 is located. 
The four sub-fields in the north-west of the image belong to the first field
(F1) observed mostly between February 2017 and January 2018, 
while those in the south-east belong to the second field (F2) observed 
between February 2018 and June 2019. 
As a result, the effective size of the field that we observed 
for NGC~2997 is about 7 $\square\degr$, 
with the 1-$\square\degr$ sub-field directly surrounding the central galaxy
observed almost 2$\times$ longer than the other sub-fields.
We stack 682, 716, and 798 individual frames of 60-second exposures
whose seeing is better than 2\arcsec\ to create the deep stacked images 
for F1 in the $BVI$-bands using {\it SWARP} \citep{2002SWARP}.
For F2, we stack 761 ($B$), 810 ($V$), and 880 ($I$) frames.
The final seeing in the stacked images is in 
the range of 1\farcs3--1\farcs6 for the $BVI$-bands
of all eight sub-fields, as shown in Table~\ref{tab:seeing}.
The average point source 3-$\sigma$ limiting magnitudes 
in the stacked images of the first field are 
$B$ = 25.14 $\pm$ 0.18 mag,
$V$ = 24.69 $\pm$ 0.24 mag,
and $I$ = 24.40 $\pm$ 0.36 mag,
while those of the second field are 
$B$ = 25.16 $\pm$ 0.24 mag,
$V$ = 24.79 $\pm$ 0.31 mag
and $I$ = 24.51 $\pm$ 0.36 mag.

We carry out photometric calibration of KMTNet CCD images using 
standard reference stars from the AAVSO Photometric All-Sky Survey (APASS)
\footnote{\url{https://www.aavso.org/apass}} catalog that are in our fields, 
and the SuperNova Analysis Package (SNAP) \footnote{\url{https://github.com/niyuanqi/SNAP}}.
In order to secure high signal-to-noise (S/N) ratios 
while avoiding CCD saturation and non-linearity effects, 
only the reference stars whose apparent magnitudes
are in the range of 15–17 mag are used in the calibration.
The APASS photometric system is based on 
the standard Johnson $BV$ and the Sloan $i$ filter bands.
It has been known that the calibration of KMTNet instrumental magnitudes
against the Johnson $B$-band magnitudes requires a color correction of
$\simeq$ 0.27 $\times$ (\bv) \citep[see][for details]{2017Park}.
After confirming the presence of the same color dependence in our images, 
we made the same color corrections on $B$-band magnitudes of the DGCs.
No such color dependency or offset has been identified for the $V$-band.
Although no color dependency exists between 
the Sloan $i$-band and the KMTNet $I$-band, 
a 0.4 mag offset is found between them from the 
comparison of the Landolt and OGLE standard stars
\citep[][see \S3.4 therein, and also Kim et al. in preparation]{2017Park}.
We, therefore, adopt 0.4 magnitude difference between 
the KMTNet $I$-band and APASS $i$-band photometry: $I$ = $i$ -- 0.4 mag.
Table~\ref{tab:zero} contains the 
zero-point magnitude offsets of our images.
According to the Galactic extinction model of
\citet{2011Schlafly}\footnote{\url{https://ned.ipac.caltech.edu/}},
the Galactic extinction in the direction of NGC~2997 are 
\(A_{B} = 0.392\), \(A_{V} = 0.296\), and \(A_{I} = 0.163\) mag.
All the magnitudes that we provide in this study are corrected 
with these extinction values.

\newlength{\abovecaptionskip}
\begin{table}
\centering
\begin{threeparttable}[b]
\caption{Seeing of the stacked images for the eight sub-fields}
\label{tab:seeing}
\begin{tabular}{ccccc}
\hline
Sub-field & & \(B\)     & \(V\)     & \(I\)     \\
          & & (arcsecs) & (arcsecs) & (arcsecs) \\
\hline
NGC2997-F1-Q0 & & 1.57$\pm$0.07   & 1.44$\pm$0.05   & 1.36$\pm$0.04   \\
NGC2997-F1-Q1 & & 1.53$\pm$0.06   & 1.43$\pm$0.05   & 1.34$\pm$0.04   \\
NGC2997-F1-Q2 & & 1.52$\pm$0.06   & 1.42$\pm$0.04   & 1.35$\pm$0.04   \\
NGC2997-F1-Q3 & & 1.48$\pm$0.04   & 1.40$\pm$0.04   & 1.34$\pm$0.04   \\
NGC2997-F2-Q0 & & 1.60$\pm$0.06   & 1.47$\pm$0.06   & 1.36$\pm$0.05   \\
NGC2997-F2-Q1 & & 1.55$\pm$0.06   & 1.44$\pm$0.05   & 1.34$\pm$0.05   \\
NGC2997-F2-Q2 & & 1.58$\pm$0.04   & 1.46$\pm$0.04   & 1.37$\pm$0.04   \\
NGC2997-F2-Q3 & & 1.52$\pm$0.04   & 1.42$\pm$0.03   & 1.34$\pm$0.06   \\
\hline
\end{tabular}
\begin{tablenotes}
\item $\dagger$ F1 and F2 indicate the two observed fields, 
and Q0 to Q3 indicate the 4 sub-fields in each field.
\end{tablenotes}
\end{threeparttable}
\end{table}

\begin{table}
\centering
\begin{threeparttable}[b]
\caption{Zero-Point Instrumental Magnitude Offsets}
\label{tab:zero}
\begin{tabular}{ccccc}
\hline
Sub-field & & \(B\) & \(V\) & \(I\) \\
\hline
NGC2997-F1-Q0 & & 28.23   & 28.01   & 28.44   \\
NGC2997-F1-Q1 & & 28.30   & 28.00   & 28.41   \\
NGC2997-F1-Q2 & & 28.23   & 28.09   & 28.39   \\
NGC2997-F1-Q3 & & 28.29   & 28.05   & 28.37   \\
NGC2997-F2-Q0 & & 28.17   & 27.98   & 28.28   \\
NGC2997-F2-Q1 & & 28.22   & 27.93   & 28.39   \\
NGC2997-F2-Q2 & & 28.21   & 28.03   & 28.32   \\
NGC2997-F2-Q3 & & 28.23   & 28.00   & 28.35   \\
\hline
\end{tabular}
\begin{tablenotes}
\item $\dagger$ 
The 1-$\sigma$ uncertainties are 0.01 mag for all the sub-fields and filters.
\end{tablenotes}
\end{threeparttable}
\end{table}

\begin{table}
\centering
\begin{threeparttable}[b]
\caption{Previously Discovered Dwarf Galaxies and Their Properties}
\label{tab:prev}
\begin{tabular}{ccc}
\hline
Name & $B$-band Magnitudes{\rm $^a$} & Redshift ($z$) \\
\hline
ESO 434-33   &  13.19, \;     13.09   &  0.0032  \\
ESO 434-27   &  15.22, \;     15.16   &  0.0031  \\
LEDA 718707  &  17.24, \;     17.02   &  0.0034  \\
ESO 434-39   &  15.55, \;     15.38   &  0.0034  \\
ESO 434-34   &  14.06, \;     14.24   &  0.0033  \\
IC 2507      &  13.12, \;     12.93   &  0.0042  \\
ESO 434-41   &  14.69, \;     14.60   &  0.0033  \\
\hline
\end{tabular}
\begin{tablenotes}
\item \rm $^a$The previously measured magnitudes 
and the magnitudes obtained in this study (see the text).
\end{tablenotes}
\end{threeparttable}
\end{table}

\section{Dwarf Galaxy Candidates and Basic Properties}\label{sec:DGC}

\subsection{Search of Dwarf Galaxy Candidates}\label{subsec:sea}
\begin{figure*}
\center
\includegraphics[width = \textwidth]{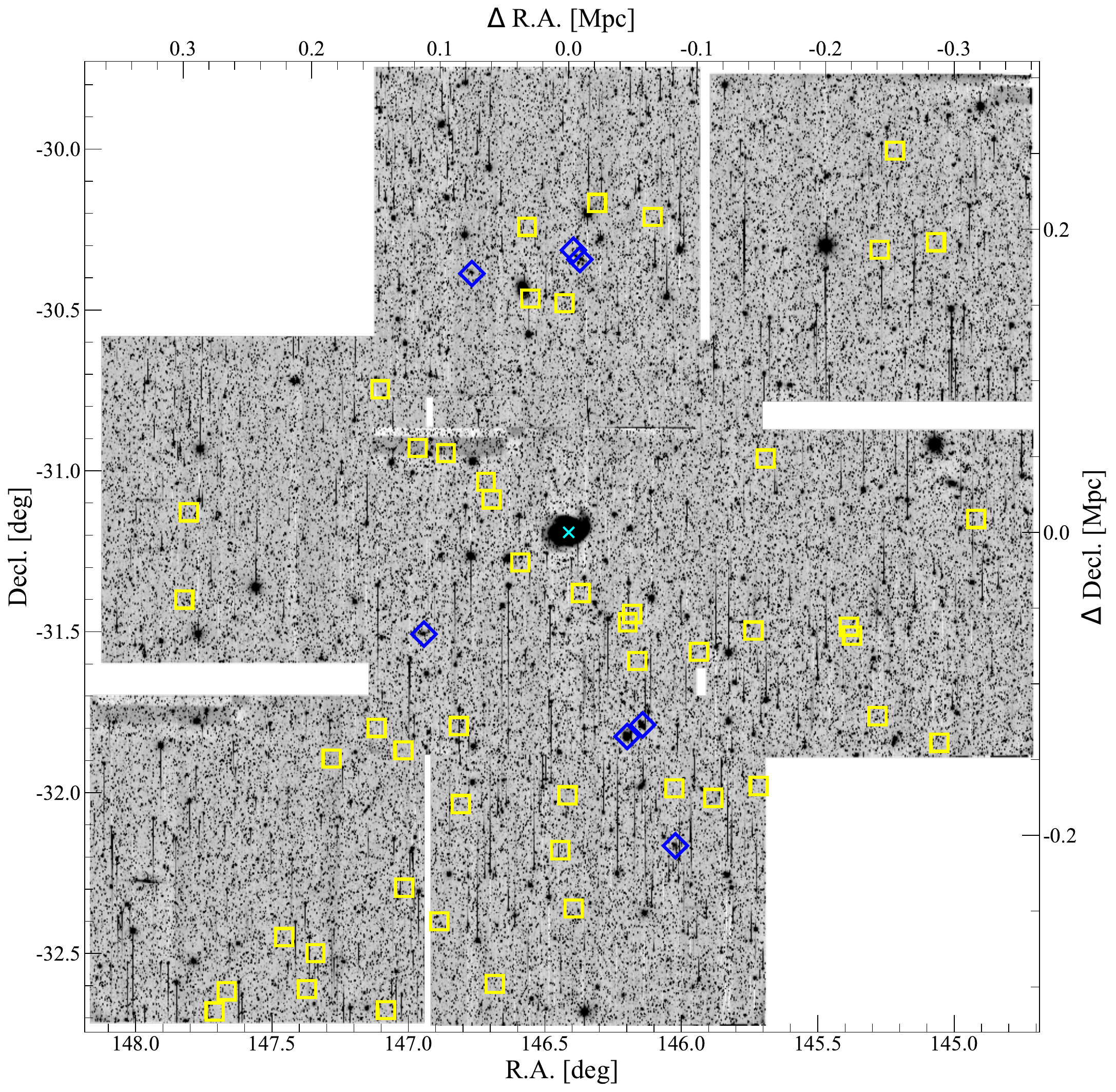}
\caption{$I$-band stacked KMTNet images of about 7$\square\degr$
around the galaxy NGC~2997 (cyan cross at the center).
The blue open diamonds mark the locations of 7 previously
discovered dwarf galaxies (See \S~\ref{subsec:sea}). 
The yellow open squares mark the other 48 DGCs.
\label{fig:MAP}}
\end{figure*}

We conduct a thorough visual inspection on stacked KMTNet images of 
the eight sub-fields to identify DGCs around NGC~2997.
We use the $I$-band images in this search of DGCs,
applying the selection criteria of being a diffuse source extending 
more than 20 pixels (or $\simeq$ 8\arcsec) that is absent of 
any prominent structures (e.g., spirals or bars), 
except for potential nucleated centers, as was done in 
\citet{2017Park} and \citet{2019Park} 
for the NGC~2784 and NGC~3585 fields, respectively.
Four\footnote{TJF, HSP, YDL and SCK from the author list} inspectors 
individually inspected the images and made a list of candidates of their own. 
44 dwarf galaxy candidates 
were co-selected by all four members in their independent identification,
and 11 more are additionally added unanimously after a second round of examination 
on candidates that were originally selected by less than 4 inspectors.
In addition to the $I$-band images, 
the $B$- and $V$-band images are also used to confirm the identified DGCs.
As a result, we identify 55 DGCs in total, with 80\% of them being selected in the first round.
Figure~\ref{fig:MAP} shows their locations around NGC~2997 
on the stacked $I$-band image. 
We note that there are more DGCs in the south-eastern part 
of NGC~2997 than in the north-western part,
especially in the region $\gtrsim$ 1.3\degr\ away from the galaxy
where there are twice as much DGCs in the south-eastern corner.
Figure~\ref{fig:cutout} shows example cutout images of 8 DGCs, 
while the images of all 55 DGCs is available in the appendix (Figure~\ref{fig:cutoutfull}).

\begin{figure*}
\center
{\includegraphics[width = \textwidth]{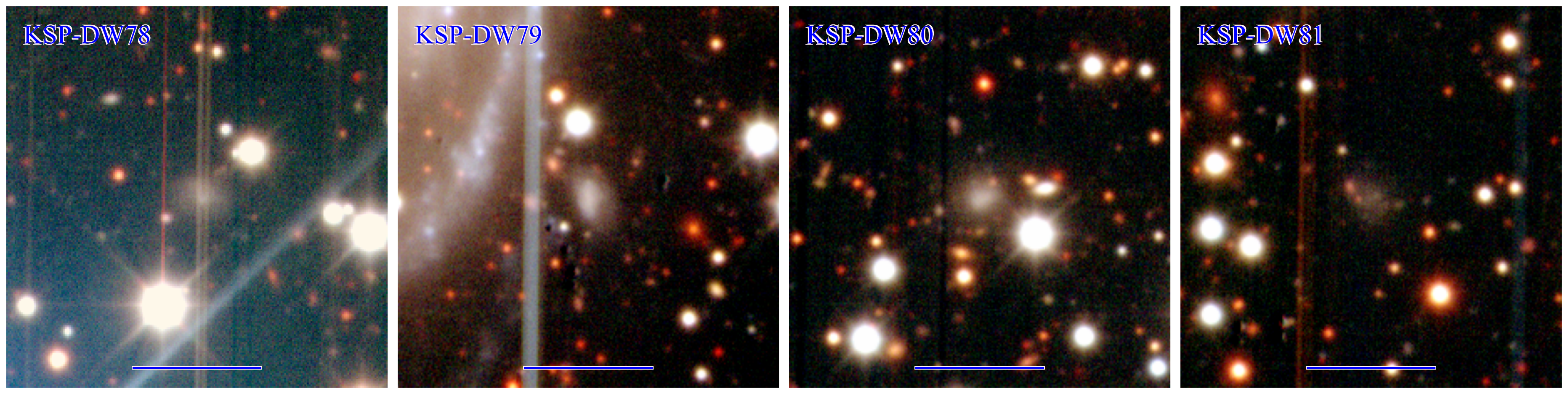}}\\
{\includegraphics[width = \textwidth]{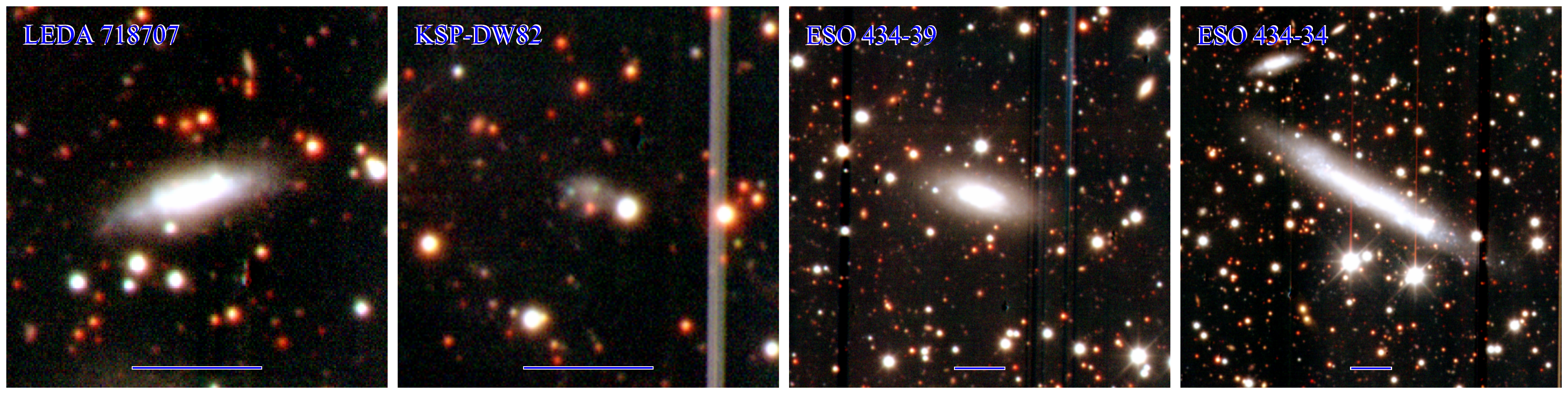}}\\
\caption{Example RGB cutout images of 8 (out of 55) DGCs in NGC~2997. 
North is up and east is to the left. 
The blue horizontal bar at the bottom of each image
corresponds to an angular size of 0\farcm5 
(= 1.77 kpc at the distance of 12.2 Mpc).
The cutout images of all 55 DGCs are available in Appendix (Figure~\ref{fig:cutoutfull}).
\label{fig:cutout}}
\end{figure*}

Out of the 55 DGCs, seven were previously identified
in the ESO and LEDA catalogs \citep{1989ESO,2003LEDA}. 
Five of these (i.e., ESO 434-27, ESO 434-31 [= IC 2507], ESO 434-33, 
ESO 434-34 and ESO 434-41) are known to be members of the loose galaxy 
group LGG~180 in which NGC~2997 is the largest member \citep{1993Garcia}.
The redshifts of the other two (i.e., ESO 434-39 and LEDA 718707)
are similar to that of NGC~2997 \citep{2009Jones}.
Although we recovered three other dwarf galaxies 
(i.e., ESO 434-30, ESO 434-43, and LEDA 100726)
that were previously presented in the catalogs, 
we exclude them from our final list because their redshifts
are significantly larger than $z$ $\simeq$ 0.0036 of NGC~2997:
0.0071 (ESO 434-30; \citealp{2004Meyer}), 
0.0083 (ESO 434-43; \citealp{1998Theureau}), 
and 0.0080 (LEDA 100726; \citealp{1995Matthews}).
Table~\ref{tab:prev} shows redshifts of the seven
previously discovered dwarf galaxies.

The two fields that we observed towards NGC~2997 are slightly different 
in the total number of stacked images because 
the second field has about 10~\% more images and is therefore 
deeper than the first by $\simlt$ 0.1 mag (Table \ref{tab:zero}). 
The photometric depth difference is at the level of the 
photometric uncertainties of the two fields, 
indicating that the difference in depth is not significant. 
This conclusion is supported by our identification of 
the exact same number of DGCs in the overlapping sub-fields 
between the first and second field: NGC2997-F1-Q2 and NGC2997-F2-Q1. 
{\tjf We conduct a completeness test of our dwarf galaxy identification
on the stacked $I$-band images by inserting 600 artificial galaxies 
created with the following seven parameters:
position angle, ellipticity, S\'ersic index, $I$-band magnitude,
effective radius, effective surface brightness, and central surface brightness.
The position angles and ellipticity of them are random distributions between
--90 and 90 degrees and 0.05 and 0.9, respectively,
while S\'ersic indices are between 0.1 and 1.,
which contain most of common cases (e.g., see bottom panel of Figure~\ref{fig:REN}).
We conduct a random sampling of the $I$-band magnitudes 
in the range of 11.3--21.7 mag, which is 1 mag broader than 
the magnitude range of the 55 DGCs in order to include dwarf galaxies 
that might be outside of our detection limits.
For the effective radius, effective surface brightness, 
and central brightness of the artificial galaxies,
we construct their Gaussian random distributions centered at the values
expected from the linear relations between them and the $I$-band magnitudes
obtained from the 55 DGCs in our sample.
The Gaussian RMS widths of these distributions are 
equivalent to 1.6 times the dispersion of the parameters 
obtained from the 55 DGCs, which effectively expands
volume of the parameter space for the artificial galaxies.
We find a 90\% completeness level down to $I$ $\simeq$ 19.0 mag 
(or M$_{V}$ $\simeq$ --10.6 mag at the distance of NGC~2997) 
and $\mu_{e,I}$ $\simeq$ 25.5 mag arcsec$^{-2}$.}
Our detection rate of bright DGCs in NGC~2997 is comparable
with that of the SAGA survey \citep{2021Mao} on 36 MW-analog galaxies.
For example, in NGC~6181, a spiral galaxy with a mass similar to that of NGC~2997,
the SAGA survey spectroscopically confirmed 9 satellites with M$_{V}$ $\lesssim$ --13 mag. 
Adopting the same satellite selection criteria of SAGA --- which are 
within the projected distance of $\sim$ 300~kpc and radial velocity $\pm$~275 km~s$^{-1}$
from the host --- for the 10 DGCs with M$_{V}$ $\lesssim$ --13 mag in our sample,
we find that 7 of them can be confirmed as satellites.
These 7 DGCs are the bright ones that were previously discovered (See Table~\ref{tab:prev}).
Since we do not have spectroscopic information for the remaining three DGCs, 
the number of confirmed satellites for NGC~2997 upon the same SAGA selection criteria
is in the range of 7--10 in our study, comparable to 9 for NGC~6181 from SAGA.

The number density of contaminating background galaxies 
in typical KMTNet stacked images was estimated to be 
1.75 $\pm$ 0.66 deg$^{-2}$ from the analysis of the KSP
images of KK~196 field \citep{2019Park},
which has been observed in very similar manner
to other KSP fields like NGC~3585 or NGC~2997. 
The distance to the KK~196 field, which belongs to the Centaurus A group, 
is only $\sim$4 Mpc, and dwarf galaxies belong to this group should appear resolved.
Applying the same visual inspection procedures 
adopted for the KSP stacked images of NGC~2784 and NGC~3585 fields,
\citet{2019Park} identified 11 dwarf galaxy candidates from the 4~deg$^{2}$ KK~196 field.
Out of these 11 identified dwarf galaxy candidates, 4 of them are resolved, 
while the rest 7 are unresolved. 
The 4 resolved candidates include 2 previously known dwarf galaxies associated
with the Centaurus group. 
The 7 unresolved candidates appear to be randomly distributed in the field.
\citet{2019Park} concluded that the 4 resolved candidates are
associated with the Centaurus group,
whereas the 7 unresolved candidates are unassociated with the group
and, therefore, marked as background sources. 
This gives the contamination level of 1.75 $\pm$ 0.66 deg$^{-2}$ 
for the stacked KSP images of KK~196 field.
Assuming the same background contamination level between the KK~196 and 
NGC~2997 field images that have very similar photometric depths, 
we estimate 7--16 as the number of potential background contamination sources 
for the 55 DGCs in the NGC~2997 group.

{\tjf We note that our background contamination estimation based on
\citet{2019Park} has not included the possibility that contamination
could be dependent on the observed properties of the DGCs. 
Results from the ELVES survey \citep{2021Carlsten} have shown that 
the background contamination rate is dependent on 
the luminosity and central surface brightness of dwarf galaxies,
such that the contaminants tend to be fainter overall but have a larger 
central surface brightness, and the number of contaminants in the ELVES survey 
starts to increase substantially for M$_{V}$ $\simgt$ --13.
This is more or less similar to the findings from \citet{2019Park}
that show all the contaminants are fainter 
than M$_{V}$ = $-$12.5 mag at the distance of NGC~2997,
indicating that the 7--16 potential contaminants in the field
are more likely to have M$_{V}$ $\simgt$ --12.5 mag
with larger central surface brightness.
Due to our visual identification procedure of DGCs (\S~\ref{subsec:sea}),}
they are less likely background early-type ellipticals,
which have steep surface brightness profiles from the center,
and are more likely blue, luminous late-type galaxies in the background.

\subsection{Surface Photometry and Catalog}\label{subsec:SBPNCAT}
\begin{figure*}
\center
{\includegraphics[width = 0.75\textwidth]{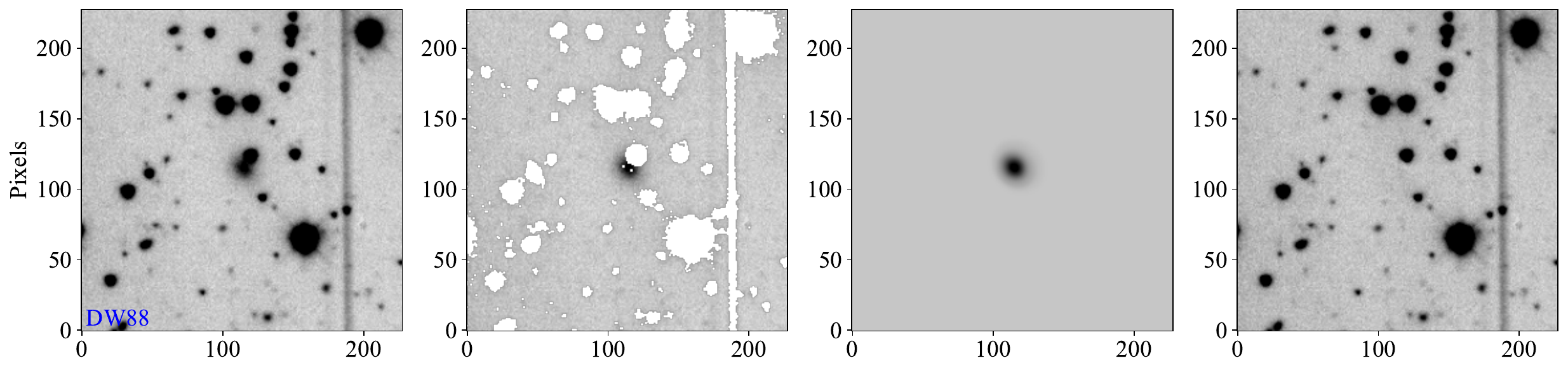}}\\
{\includegraphics[width = 0.75\textwidth]{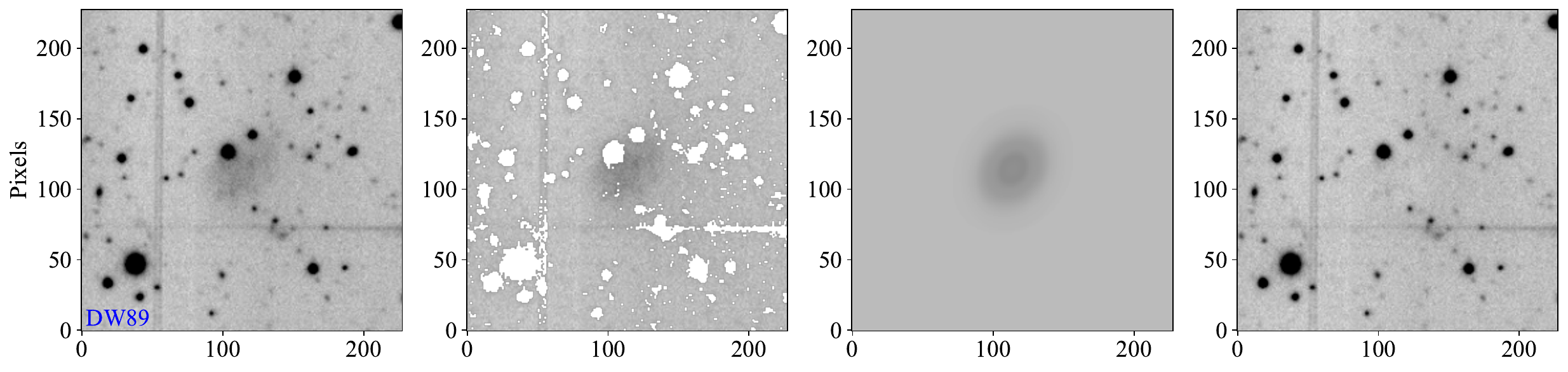}}\\
{\includegraphics[width = 0.75\textwidth]{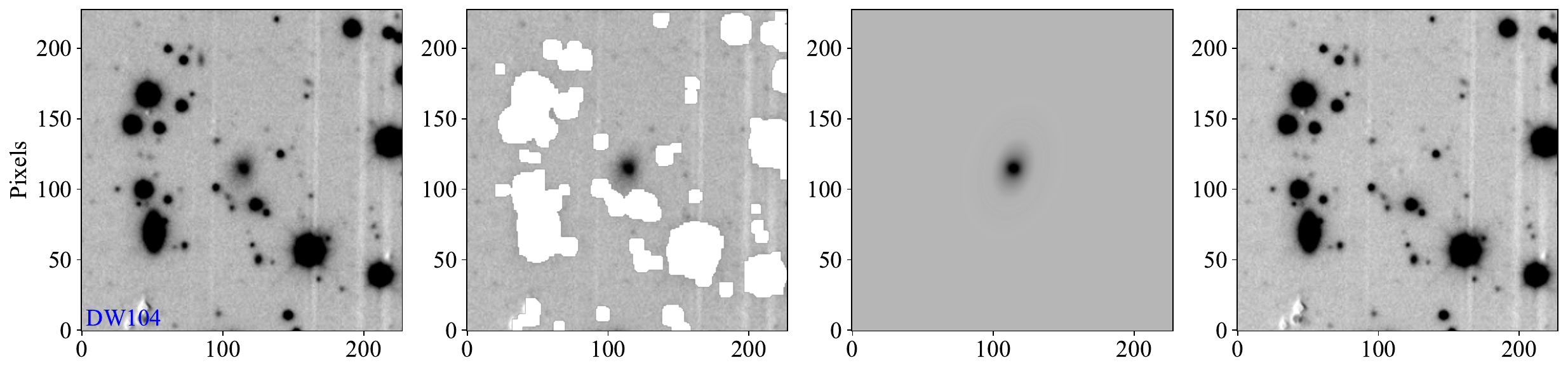}}\\
{\includegraphics[width = 0.75\textwidth]{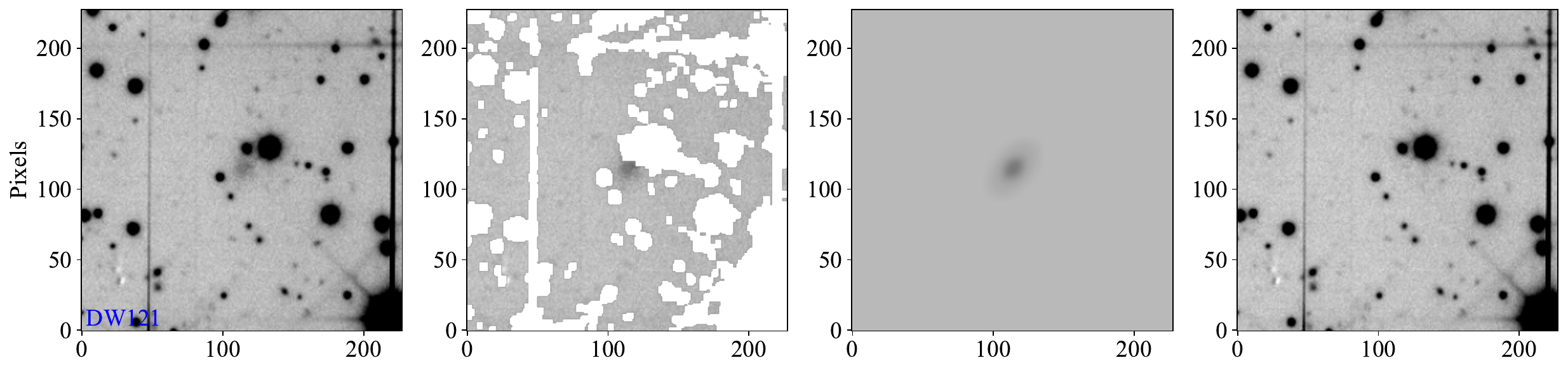}}\\
{\includegraphics[width = 0.75\textwidth]{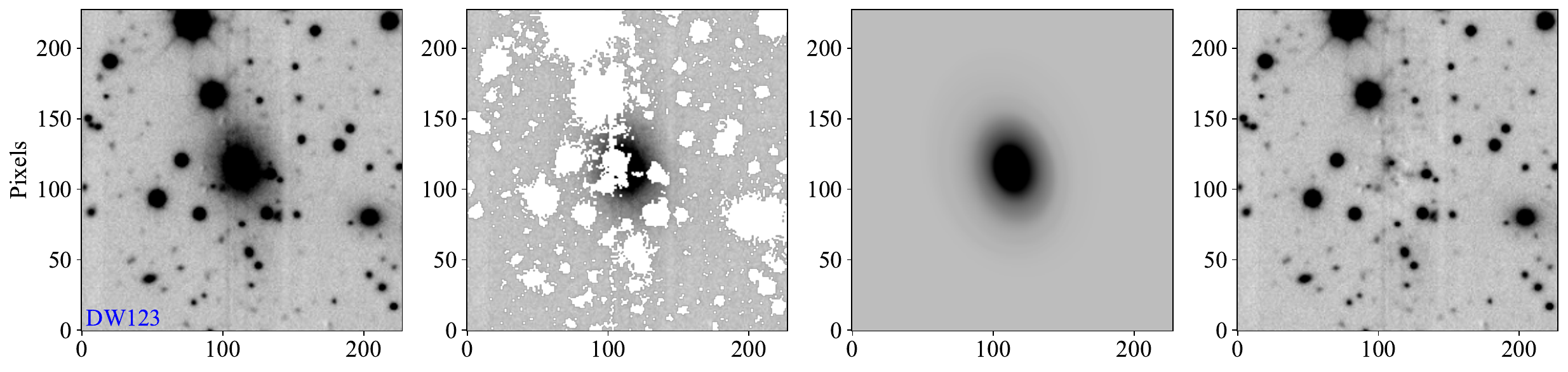}}\\
{\includegraphics[width = 0.75\textwidth]{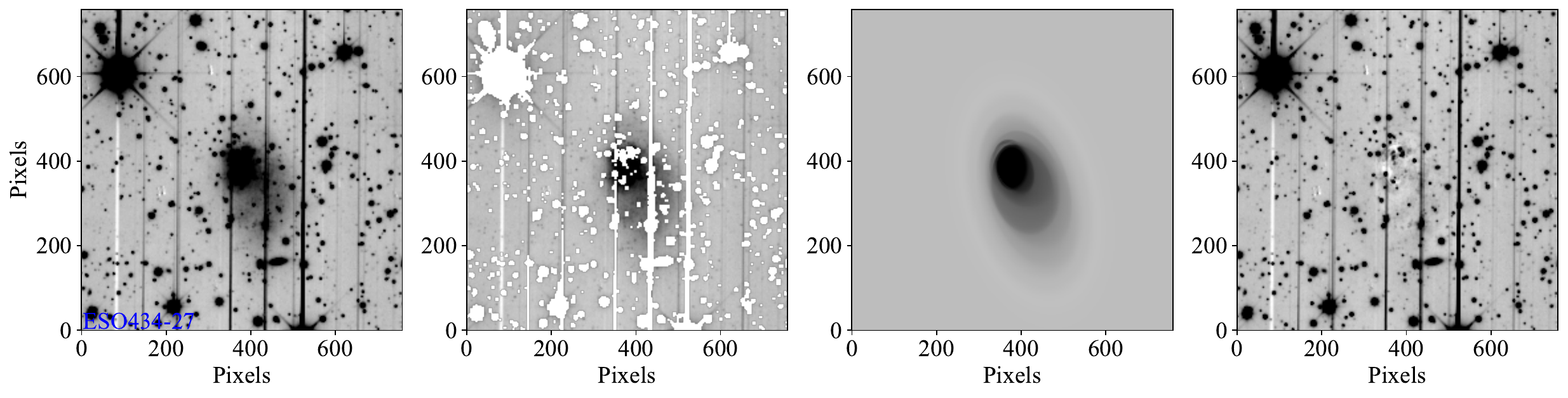}}\\
\caption{Cutout images of the six example DGCs 
demonstrating our photometric processes for obtaining isophotal models:
DW88, DW89, DW104, DW121, DW123 and ESO434-27 from top to bottom.
The rows are for the data (or stacked images; first row), 
the data with masks (second row), fitted isophotal models (third row),
and residual of the data after subtraction of the isophotal models (see \S~\ref{subsec:SBPNCAT}). 
\label{fig:DMR}}
\end{figure*}

We carry out surface photometry of the 55 DGCs by constructing 
their radial isophotal profiles using the {\it ELLIPSE} fitting task
from the python-based photometry package {\it Photutils}
\footnote{https://photutils.readthedocs.io/en/stable/}. 
The {\it ELLIPSE} task calculates the best-fit isophotal parameters
(i.e., the central position, positional angle, and ellipticity)
by fitting a galaxy based on the initial visual estimations of 
those parameters under the assumption of an elliptical geometry.
The typical linear step-size of fitting is 2--3 pixels 
along the semi-major axis; however,
for the brighter and larger ($r_{e}$ $>$ 15\arcsec) sources, 
we conduct the fitting in logarithmic scale for efficiency.
We mask out foreground/background objects 
that overlap with the DGCs when possible.
The intensity of each isophote is determined by the median value 
of the isophote's pixels, which effectively minimizes the influence
of any interfering artifacts/sources in the foreground/background;
and the total intensity of each isophote is obtained 
by multiplying the median value with the total number of pixels within. 
The sky background level, which is estimated using 
the median value of source-free pixels surrounding a DGC, 
is subtracted from the isophotal intensities.
The outermost isophote of a DGC is determined 
when the isophotal intensity is $\simeq$ 1-$\sigma$ 
above the sky background level.
We then estimate the total instrumental magnitudes of the DGCs
by integrating the total intensities of their isophotal models.
Figure~\ref{fig:DMR} shows example results for 6 dwarf galaxy candidates 
where the data, masked data, models, and residuals are shown.
For the $B$- and $V$-band images of the DGCs, 
we follow the same procedure above using the isophotal models 
from $I$-band images to construct their surface brightness profiles 
as well as their instrumental magnitudes in $B$- and $V$-bands.
Finally, we obtain the corrected apparent magnitudes of the DGCs
as outlined in \S\ref{sec:obs}, 
including the conversion from instrumental magnitudes to apparent magnitudes 
using zero-point offsets (Table~\ref{tab:zero}),  
$B$-band color correction and $I$-band magnitude offset correction, 
as well as Galactic extinction correction.

{\tjf We investigate the effective performance of our isophotal photometry by comparing} 
the isophotal model magnitudes of 13 relatively isolated DGCs in our sample
with their magnitudes obtained by direct aperture photometry.
The 13 DGCs are the entirety of DGCs that appear to be 
free from apparently significant interference by overlapping sources.
The magnitude differences between the two methods are small, 
with 75\% of them showing differences $\simlt$ 0.05 mag, 
although a few of them reach differences of $\sim$ 0.2 mag.
We note that the isophotal method produces slightly larger (= fainter) magnitudes
than the aperture photometry method.
One possible explanation for this is the contribution by 
faint overlapping sources to the aperture photometry. 
The 7 previously-discovered dwarf galaxies in Table~\ref{tab:prev},
for which $B$-band magnitudes from previous studies are available,
do not belong to this group of 13 apparently isolated DGCs.
The average $B$-band magnitude differences between their literature values
and isophotal model magnitudes obtained in this study is 0.14 mag.
We, therefore, conclude that there exists a reasonable agreement between
magnitudes obtained by our isophotal method and aperture photometry method,
as well as those obtained in previous studies.

{\tjf All the DGCs in our sample have $r_{e}$ $\geq$ 2\farcs39,
which is greater than the seeing of 
our stacked images in the range of 1\farcs3--1\farcs6 (Table~\ref{tab:seeing}), 
and the PSF effects to our photometry of the DGCs are expected to be small.
We examine the PSF effects to our photometry of eight 
(= DW-93, 104, 88, 102, 115, 101, 120, 118) smallest DGCs in our sample with 
$r_{e}$ $<$ 4\arcsec\ given that the PSF effects can be more apparent for small DGCs.
First, the differences between the total $BVI$ magnitudes 
obtained by our isophotal models and by PSF convoluted models
of these eight DGCs are very small, with most of them being $\le$ 0.01 magnitudes.
We also find that the average differences of their estimated effective radius, 
effective brightness, and S\'ersic index between the two methods  
are 0\farcs15, 0\farcs07 mag arcsec$^{-2}$, and 0.03, respectively.
These values are much smaller than their propagated measurement uncertainties 
of 0\farcs94, 0.52 mag arcsecs$^{-2}$, and 0.2, respectively,
based on their uncertainties in Table~\ref{tab:cat}.
We therefore conclude that the PSF effect on our galaxy parameters estimation is not significant.}

\begin{figure*}
\center
{\includegraphics[width = \textwidth]{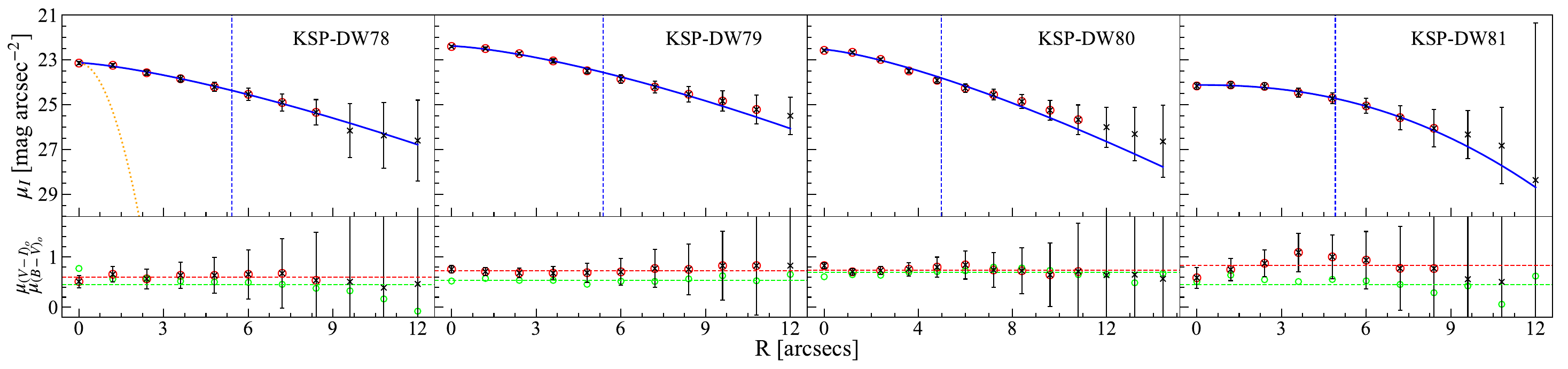}}\\
{\includegraphics[width = \textwidth]{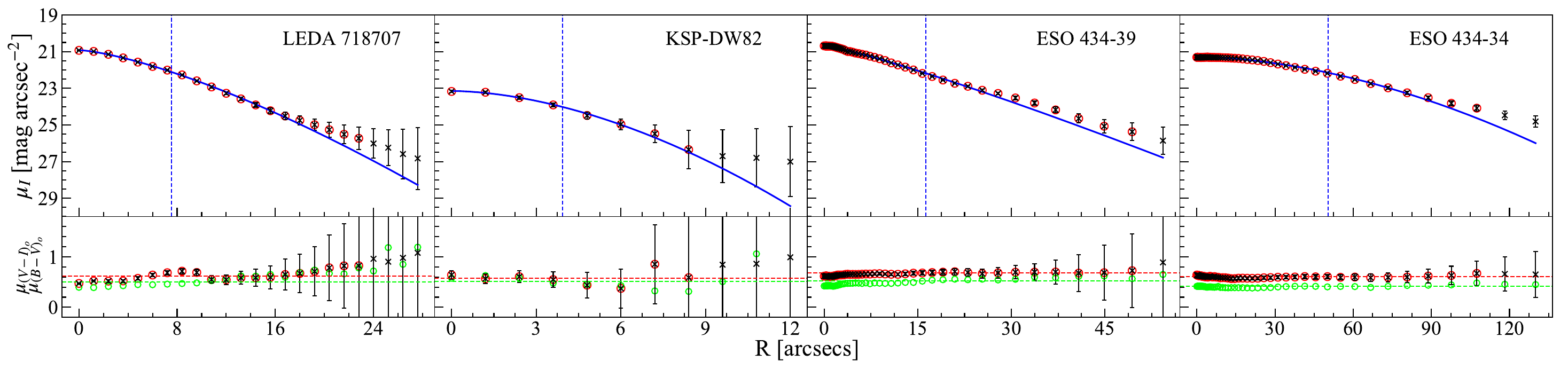}}\\
\caption{Eight examples of surface brightness profiles of the DGCs. 
{\it Upper Panels}: $I$-band surface brightness profiles.
The black crosses with error bars, often encircled by a red circle, 
represent $\mu_I$ and its uncertainty. 
Only those data points with a red circle are used in 1D S\'ersic fitting.
The blue solid curves show the resulting best-fit S\'ersic profiles. 
The vertical blue dashed lines mark the estimated effective radii. 
{\it Lower Panels}: \bvo\ and \vio\ color profiles are shown in 
green circles and black crosses, respectively. 
The green and red horizontal dashed lines represent the mean
\bvo\ and \vio\ colors obtained from difference in total apparent magnitudes
(See \S~\ref{subsec:SBPNCAT}), respectively.
{\tjf The dotted orange curve in the surface brightness profile of 
KSP-DW78 represents the PSF profile of the image to which the dwarf galaxy belongs.
The surface brightness profiles of all 55 DGCs and the PSF profiles of all 8 images
are available in Appendix (Figure~\ref{fig:SBPFULL}).}
\label{fig:SBP}}
\end{figure*}

Figure~\ref{fig:SBP} shows eight examples of surface brightness profiles 
obtained from our surface photometry (full figure in appendix, Figure~\ref{fig:SBPFULL}).
The top panels show the $I$-band surface brightness profiles (black crosses)
that we obtained using median values of their isophotes for the 55 DGCs. 
To examine the radial distributions of light, 
we fit the surface brightness profiles with the S\'ersic function 
$\mu_I$ = $\mu_{o,I}$ + 1.0857 $(r/r_{o,I})^{1/n_{I}}$, 
where $\mu_I$, $\mu_{o,I}$, $r$, $r_{o,I}$ and $n_{I}$ 
are the surface brightness, central surface brightness, 
radius, scale length, and S\'ersic index. 
During the fitting, we only use isophotes with intensities
higher than the sky background level at a confidence level of 90\% 
(1.6-$\sigma$), and those isophotes used in the fitting
are marked by red circles in Figure~\ref{fig:SBP}.
The solid blue curves in the figure represent the best-fit S\'ersic profiles,
and the vertical dashed lines in blue indicate the effective radius
$r_{e}$ = $(2.3026 \times Bn)^{n}$ $r_{o}$, 
where $Bn$ = 0.868 $n$ -- 0.142 \citep{2009Chiboucas}. 
{\tjf The orange dotted curves in 8 of the surface brightness profiles 
are the PSF models of the images to which the corresponding DGCs belong.}
The bottom panels show the DGCs' color profiles --- 
green open circles for \bvo\ and black crosses for \vio.
The green and red dashed lines indicate the values for \bvo\ and \vio\ colors obtained 
using the difference in total magnitudes, respectively.
Table~\ref{tab:cat} lists the parameters of the 55 DGCs including their names, coordinates, 
$I$-band total apparent magnitudes ($I$), \vio\ and \bvo\ colors, three best-fit S\'ersic parameters 
(i.e., $\mu_{o,I}$, $r_{o,I}$, $n_{I}$), effective radius in $I$-band ($r_{e,I}$), 
and $V$-band total absolute magnitudes (M$_{V}$). 
The final column (``Comments'') shows if the candidates are morphologically identified
as ultra-diffuse, or nucleated, or irregular galaxies (see \S~\ref{subsec:NUG}). 
All the DGCs are brighter than 21.24 mag in $I$-band, 
whereas the faintest $B$-band and $V$-band magnitudes are 23.98 and 22.41 mag, respectively.
Their colors are in the range of 0.39 -- 1.17 mag for \vio, and 0.25 -- 1.57 mag for \bvo\ .
The central surface brightnesses of all candidates  
are brighter than $\mu_{o,B}$ $\simeq$ 27.9, $\mu_{o,V}$ $\simeq$ 26.7, 
and $\mu_{o,I}$ $\simeq$ 25.4 mag arcsec$^{-2}$.
The $I$-band S\'ersic scale lengths and the curvature indices 
vary in the range of 2\farcs5 $\lesssim$ $r_{o,I}$ $\lesssim$ 57\farcs9\
and 0.25 $\lesssim$ $n_{I}$ $\lesssim$ 1.39, respectively. 
The $I$-band effective radii ($r_{e,I}$) are in the range of 2\farcs4--50\farcs4, 
or 0.142--2.97 kpc at the distance of 12.2 Mpc.
About 87\% (or 48 out of 55) have effective radii smaller than 12\arcsec\ (or 0.71 kpc).
The largest newly discovered DGC in our sample is KSP-DW106, 
a very elongated source with an $I$-band effective radius of 22\farcs9.
However, it is smaller than most of the 7 previously discovered candidates.

\begin{table*}
\scriptsize
\begin{threeparttable}
\caption{Catalog of NGC~2997 Dwarf Galaxy Candidates}
\label{tab:cat}
\begin{tabular}{cccccccccccc}
\hline
ID & RA         & DEC        & $I$   & \vio  & \bvo  & \(\mu_{o,I}\)       & \(r_{o,I}\) & \(n_{I}\) & \(r_{e,I}\) & M$_{V}$ & Comments\rm $\dagger$\\
   & (hh:mm:ss) & (dd:mm:ss) & (mag) & (mag) & (mag) & (mag arcsec$^{-2}$) & (arcsec)    &           & (arcsec)    & (mag)   &                      \\
\hline
KSP-DW78    & 9:45:14.2 & -30:10:01.2 & 18.44 & 0.60 & 0.45 & 23.13$\pm$0.07 & 4.92$\pm$0.42  & 0.73$\pm$0.14 & 5.42  & -11.39 &     \\
KSP-DW79    & 9:46:12.8 & -30:27:51.2 & 17.98 & 0.72 & 0.53 & 22.39$\pm$0.04 & 5.07$\pm$0.27  & 0.71$\pm$0.07 & 5.36  & -11.72 & dIr \\
KSP-DW80    & 9:45:42.8 & -30:28:42.9 & 17.82 & 0.73 & 0.69 & 22.54$\pm$0.04 & 4.43$\pm$0.21  & 0.75$\pm$0.06 & 4.98  & -11.88 &     \\
KSP-DW81    & 9:46:15.7 & -30:14:29.6 & 19.25 & 0.83 & 0.45 & 24.12$\pm$0.11 & 6.37$\pm$0.62  & 0.44$\pm$0.16 & 4.91  & -10.36 & dIr \\
LEDA 718707 & 9:45:34.9 & -30:18:51.3 & 15.90 & 0.62 & 0.50 & 20.92$\pm$0.01 & 7.05$\pm$0.06  & 0.71$\pm$0.01 & 7.54  & -13.91 & dIr \\
KSP-DW82    & 9:44:25.2 & -30:12:38.3 & 18.92 & 0.58 & 0.51 & 23.15$\pm$0.05 & 4.45$\pm$0.24  & 0.56$\pm$0.07 & 3.93  & -10.93 &     \\
ESO 434-39  & 9:47:04.2 & -30:23:12.5 & 14.17 & 0.68 & 0.53 & 20.66$\pm$0.00 & 12.22$\pm$0.05 & 0.86$\pm$0.01 & 16.30 & -15.58 & dIr \\
ESO 434-34  & 9:45:29.3 & -30:20:35.5 & 13.22 & 0.60 & 0.42 & 21.31$\pm$0.00 & 57.87$\pm$0.36 & 0.55$\pm$0.01 & 50.40 & -16.61 & dIr \\
KSP-DW83    & 9:41:05.6 & -30:18:45.3 & 17.23 & 0.71 & 0.68 & 21.61$\pm$0.02 & 3.33$\pm$0.11  & 1.15$\pm$0.05 & 7.32  & -12.49 & dIr \\
KSP-DW84    & 9:40:16.3 & -30:17:22.7 & 17.94 & 0.48 & 0.45 & 23.86$\pm$0.08 & 10.31$\pm$0.97 & 0.46$\pm$0.17 & 8.13  & -12.0  & dIr \\
KSP-DW85    & 9:40:52.1 & -30:00:17.5 & 19.10 & 0.87 & 1.10 & 23.48$\pm$0.08 & 4.49$\pm$0.35  & 0.49$\pm$0.14 & 3.65  & -10.46 &     \\
KSP-DW86    & 9:48:04.3 & -31:52:07.1 & 17.06 & 0.71 & 0.29 & 22.58$\pm$0.05 & 7.13$\pm$0.44  & 0.82$\pm$0.09 & 8.92  & -12.66 & dIr \\
KSP-DW87    & 9:47:15.7 & -31:47:34.7 & 19.90 & 0.83 & 0.26 & 23.60$\pm$0.09 & 3.81$\pm$0.35  & 0.54$\pm$0.15 & 3.26  & -9.71  &     \\
KSP-DW88    & 9:48:27.9 & -31:47:56.3 & 18.89 & 0.85 & 0.68 & 22.61$\pm$0.04 & 3.12$\pm$0.13  & 0.69$\pm$0.06 & 3.21  & -10.69 &     \\
KSP-DW89    & 9:44:38.9 & -31:35:28.2 & 18.30 & 0.59 & 0.25 & 24.47$\pm$0.11 & 11.48$\pm$1.12 & 0.36$\pm$0.17 & 8.21  & -11.54 &     \\
IC 2507     & 9:44:34.0 & -31:47:22.4 & 11.82 & 0.78 & 0.33 & 19.55$\pm$0.00 & 23.98$\pm$0.13 & 0.66$\pm$0.01 & 23.71 & -17.83 & dIr \\
KSP-DW90    & 9:44:47.2 & -31:28:14.5 & 19.70 & 0.84 & 0.65 & 23.32$\pm$0.06 & 3.67$\pm$0.21  & 0.48$\pm$0.09 & 2.94  & -9.89  &     \\
KSP-DW91    & 9:45:28.3 & -31:22:47.4 & 17.88 & 0.87 & 0.67 & 22.25$\pm$0.03 & 2.76$\pm$0.19  & 1.39$\pm$0.12 & 9.65  & -11.68 & N   \\
KSP-DW92    & 9:46:21.7 & -31:17:03.4 & 18.56 & 0.85 & 0.71 & 24.10$\pm$0.21 & 9.71$\pm$2.90  & 0.70$\pm$0.58 & 10.19 & -11.02 &     \\
KSP-DW93    & 9:46:46.9 & -31:05:20.2 & 20.62 & 1.08 & 0.60 & 24.51$\pm$0.16 & 4.56$\pm$0.74  & 0.52$\pm$0.27 & 3.83  & -8.74  &     \\
KSP-DW94    & 9:46:51.8 & -31:02:02.0 & 21.24 & 1.17 & 1.57 & 25.14$\pm$0.33 & 5.69$\pm$2.77  & 0.65$\pm$0.89 & 5.61  & -8.02  &     \\
KSP-DW95    & 9:47:51.8 & -30:55:44.2 & 20.00 & 0.99 & 0.55 & 23.44$\pm$0.07 & 2.54$\pm$0.25  & 0.78$\pm$0.15 & 3.00  & -9.44  &     \\
KSP-DW96    & 9:47:27.1 & -30:56:40.4 & 16.85 & 0.89 & 0.58 & 21.87$\pm$0.03 & 5.24$\pm$0.17  & 0.64$\pm$0.04 & 5.05  & -12.69 &     \\
ESO 434-41  & 9:47:46.4 & -31:30:27.6 & 13.57 & 0.68 & 0.34 & 21.86$\pm$0.00 & 33.60$\pm$0.45 & 0.82$\pm$0.02 & 41.86 & -16.18 & dIr \\
KSP-DW97    & 9:40:12.9 & -31:50:42.3 & 17.72 & 0.58 & 0.52 & 22.27$\pm$0.03 & 3.76$\pm$0.17  & 0.78$\pm$0.06 & 4.44  & -12.13 &     \\
KSP-DW98    & 9:41:07.5 & -31:45:45.1 & 19.51 & 0.64 & 0.65 & 24.22$\pm$0.16 & 5.60$\pm$0.85  & 0.41$\pm$0.22 & 4.18  & -10.29 &     \\
KSP-DW99    & 9:41:32.8 & -31:29:04.6 & 19.98 & 0.51 & 0.42 & 23.27$\pm$0.06 & 3.15$\pm$0.23  & 0.69$\pm$0.10 & 3.26  & -9.94  & dIr \\
KSP-DW100   & 9:41:29.6 & -31:30:43.3 & 16.92 & 0.89 & 0.75 & 21.54$\pm$0.01 & 3.80$\pm$0.08  & 0.96$\pm$0.03 & 5.90  & -12.63 &     \\
KSP-DW101   & 9:42:56.4 & -31:29:46.9 & 20.83 & 0.47 & 0.83 & 24.71$\pm$0.29 & 3.73$\pm$1.20  & 0.46$\pm$0.58 & 2.93  & -9.14  &     \\
KSP-DW102   & 9:39:40.8 & -31:08:57.6 & 20.83 & 1.06 & 1.38 & 23.77$\pm$0.10 & 3.39$\pm$0.37  & 0.59$\pm$0.18 & 3.09  & -8.54  &     \\
KSP-DW103   & 9:42:45.7 & -30:57:42.3 & 18.64 & 0.77 & 0.51 & 22.31$\pm$0.03 & 2.63$\pm$0.10  & 0.87$\pm$0.05 & 3.57  & -11.03 &     \\
KSP-DW104   & 9:51:17.2 & -31:23:59.5 & 18.91 & 0.74 & 0.41 & 22.48$\pm$0.03 & 2.73$\pm$0.13  & 0.88$\pm$0.06 & 3.72  & -10.78 &     \\
KSP-DW105   & 9:51:13.1 & -31:07:44.3 & 17.37 & 0.82 & 0.49 & 21.33$\pm$0.01 & 2.71$\pm$0.06  & 1.10$\pm$0.03 & 5.38  & -12.25 & dIr \\
KSP-DW106   & 9:48:24.8 & -30:44:47.3 & 17.68 & 0.63 & 0.57 & 24.59$\pm$0.11 & 25.40$\pm$2.60 & 0.58$\pm$0.18 & 22.86 & -12.12 & U   \\
KSP-DW107   & 9:43:44.8 & -31:33:43.0 & 18.56 & 0.39 & 0.85 & 25.36$\pm$0.20 & 16.53$\pm$2.58 & 0.25$\pm$0.25 & 10.65 & -11.48 &     \\
KSP-DW108   & 9:44:43.4 & -31:26:40.4 & 16.45 & 0.69 & 0.61 & 21.22$\pm$0.01 & 3.41$\pm$0.07  & 1.19$\pm$0.02 & 8.04  & -13.29 & dIr \\
KSP-DW109   & 9:50:50.4 & -32:40:48.0 & 19.07 & 0.91 & 0.47 & 23.68$\pm$0.08 & 5.39$\pm$0.47  & 0.61$\pm$0.14 & 5.03  & -10.45 &     \\
KSP-DW110   & 9:48:19.6 & -32:40:33.9 & 17.87 & 0.82 & 0.49 & 23.23$\pm$0.04 & 7.13$\pm$0.31  & 0.52$\pm$0.07 & 5.98  & -11.73 &     \\
KSP-DW111   & 9:49:21.7 & -32:29:56.1 & 19.25 & 0.89 & 0.62 & 23.49$\pm$0.07 & 4.75$\pm$0.35  & 0.66$\pm$0.11 & 4.70  & -10.29 &     \\
KSP-DW112   & 9:49:29.2 & -32:36:38.8 & 19.10 & 0.90 & 0.82 & 23.34$\pm$0.06 & 4.50$\pm$0.29  & 0.63$\pm$0.09 & 4.32  & -10.44 &     \\
KSP-DW113   & 9:50:39.9 & -32:37:03.2 & 19.77 & 0.93 & 0.47 & 23.49$\pm$0.06 & 3.62$\pm$0.24  & 0.53$\pm$0.11 & 3.08  & -9.73  &     \\
KSP-DW114   & 9:48:03.9 & -32:17:44.1 & 19.30 & 0.72 & 0.42 & 23.58$\pm$0.06 & 5.05$\pm$0.29  & 0.55$\pm$0.09 & 4.37  & -10.41 &     \\
KSP-DW115   & 9:49:07.5 & -31:53:39.3 & 19.90 & 0.61 & 0.37 & 23.52$\pm$0.07 & 3.22$\pm$0.27  & 0.61$\pm$0.12 & 3.01  & -9.92  &     \\
KSP-DW116   & 9:49:49.3 & -32:26:55.5 & 15.86 & 0.96 & 0.64 & 21.78$\pm$0.02 & 8.04$\pm$0.18  & 0.86$\pm$0.03 & 10.75 & -13.61 &     \\
KSP-DW117   & 9:45:34.7 & -32:21:37.1 & 19.03 & 0.85 & 0.58 & 23.48$\pm$0.06 & 4.85$\pm$0.33  & 0.53$\pm$0.09 & 4.10  & -10.55 &     \\
KSP-DW118   & 9:47:33.0 & -32:24:00.6 & 20.28 & 0.89 & 0.71 & 23.42$\pm$0.09 & 2.61$\pm$0.27  & 0.59$\pm$0.15 & 2.39  & -9.25  &     \\
KSP-DW119   & 9:45:40.3 & -32:00:30.7 & 17.90 & 0.92 & 0.45 & 22.93$\pm$0.03 & 8.80$\pm$0.28  & 0.45$\pm$0.05 & 6.87  & -11.61 & dIr \\
KSP-DW120   & 9:44:06.2 & -31:59:11.5 & 19.54 & 0.80 & 0.82 & 22.97$\pm$0.04 & 2.59$\pm$0.15  & 0.75$\pm$0.07 & 2.91  & -10.09 &     \\
KSP-DW121   & 9:43:31.9 & -32:00:59.2 & 19.80 & 0.88 & 0.79 & 24.01$\pm$0.12 & 4.67$\pm$0.63  & 0.67$\pm$0.19 & 4.69  & -9.75  &     \\
KSP-DW122   & 9:45:46.7 & -32:10:44.0 & 19.18 & 0.73 & 0.64 & 22.79$\pm$0.04 & 3.68$\pm$0.21  & 0.81$\pm$0.08 & 4.49  & -10.53 & dIr \\
KSP-DW123   & 9:47:13.9 & -32:02:07.7 & 16.26 & 0.84 & 0.61 & 21.98$\pm$0.02 & 7.84$\pm$0.17  & 0.72$\pm$0.03 & 8.47  & -13.33 &     \\
KSP-DW124   & 9:46:44.0 & -32:35:41.6 & 17.24 & 0.82 & 0.66 & 22.52$\pm$0.03 & 7.18$\pm$0.31  & 0.86$\pm$0.06 & 9.55  & -12.38 &     \\
ESO 434-33  & 9:44:47.6 & -31:49:32.0 & 12.08 & 0.66 & 0.35 & 21.52$\pm$0.01 & 49.43$\pm$0.54 & 0.41$\pm$0.01 & 36.99 & -17.69 & dIr \\
ESO 434-27  & 9:44:05.2 & -32:09:54.3 & 13.86 & 0.75 & 0.55 & 22.06$\pm$0.01 & 25.50$\pm$0.43 & 0.86$\pm$0.02 & 33.76 & -15.83 & dIr \\
KSP-DW125   & 9:42:51.9 & -31:58:46.2 & 19.82 & 0.73 & 0.81 & 23.17$\pm$0.06 & 2.65$\pm$0.22  & 0.79$\pm$0.12 & 3.19  & -9.88  &     \\
\hline
\end{tabular}
\begin{tablenotes}
\item $\dagger$ Candidates identified as UDG (U), Nucleated (N), or irregularly shaped (dIr).
\item $I$ and M$_V$ are apparent $I$-band magnitude 
and absolute $V$-band magnitude, respectively,
estimated using the fitted parameters of the S\'ersic function. 
$\mu_{o,I}$, $r_{o,I}$, and $n_{I}$ are the central surface brightness, 
scale length, and the S\'ersic index 
from the fitting of $I$-band images, respectively. 
$r_{e,I}$ is the effective radius obtained from $r_{o,I}$ and $n_{I}$.
\vio\ and \bvo\ are the observed mean colors after correction 
of the reddening in the Milky Way.
\end{tablenotes}
\end{threeparttable}
\end{table*}

\subsection{Morphological Classifications: 
Nucleated, Ultra-Diffuse, and Irregular Dwarf Galaxies}\label{subsec:NUG}
Dwarf galaxies with distinctively bright and compact cores have been 
morphologically classified as nucleated dwarf galaxies
\citep[e.g.,][]{2006Cote,2009TrenthamTully,2019Park,2021Zaritsky}.
We identify KSP-DW91 to be the only nucleated dwarf galaxy
out of the 55 DGCs in NGC~2997 based on the presence of 
a clearly distinguishable central point-source core
with color similar to the rest of the galaxy (Figure~\ref{fig:cutoutfull}),
although 0.22 mag arcsec$^{-2}$ difference of the core in the $V$-band 
from the S\'ersic-fit expected brightness is a bit smaller than 
those found in nucleated dwarf galaxies in other groups
\citep[e.g.,][]{2019Park}.
The lack of nucleated dwarf galaxies in NGC~2997 
is in stark contrast with the range of 10--30 \% of the incidence rate from previous studies 
\citep[e.g.,][]{2009TrenthamTully,2017Park,2019Park}.

Dwarf galaxies with extremely diffuse morphological features have been
classified as ultra-diffuse galaxies 
\citep[UDGs; e.g.,][]{2015vanDokkum,2018Muller,2019Park}.
For example, \citet{2015vanDokkum} applies the following criteria
to identify UDGs: $\mu_{o,V}$ $\gtrsim$ 23.7 mag arcsec$^{-2}$ 
and $r_{e,V}$ $\gtrsim$ 1.5 kpc.
The two DGCs with the most extreme diffuse distribution of 
surface brightness in our sample are 
KSP-DW106 ($r_{e,V}$ = 1.38 $\pm$ 0.4 kpc, 
$\mu_{o,V}$ = 25.23 $\pm$ 0.11 mag arcsec$^{-2}$)
and KSP-DW107 ($r_{e,V}$ = 0.66 $\pm$ 0.48 kpc, 
$\mu_{o,V}$ = 25.78 $\pm$ 0.17 mag arcsec$^{-2}$).
If we apply the same criteria above, KSP-DW106 can be classified 
as a potential UDG, given its effective radius uncertainty. 

Finally, we identify 17 DGCs irregularly-shaped dwarf galaxies 
in our sample based on the presence of 
clearly non-elliptical morphology,
faint but still identifiable structures,
or the peak emission being off the geometrical center.
Their average values of effective radius ($I$-band),
absolute magnitudes ($V$-band), 
and \bvo\ color are 16\farcs07 (or 0.95 kpc), 
--13.56 mag, and 0.47 mag, respectively.
These irregularly-shaped dwarf galaxies,
which include all the 7 previously discovered dwarf galaxies
(Table~\ref{tab:cat}),
appear to be significantly larger, brighter, and bluer
than r$_{e,I}$ $\simeq$ 8\farcs80 (or 0.52 kpc), 
M$_{V}$ $\simeq$ --11.61 mag, and \bvo\ $\simeq$ 0.60 mag,
which are the average values for the entire DGC population.
We note that it is easier to select large DGCs as irregular 
than smaller ones due to our selection criteria, 
and there could be a group of small irregularly-shaped DGCs that was not selected. 
The real difference in size, brightness and color between
the irregularly-shaped DGCs and the entire DGC population
of NGC~2997, therefore, could be smaller.

\section{Properties of Dwarf Galaxy Candidates in the NGC~2997 group}\label{sec:group}

\subsection{Radial Distribution of Number Density and Color}\label{subsec:RND}
\begin{figure}
\center
\includegraphics[width=0.481\textwidth]{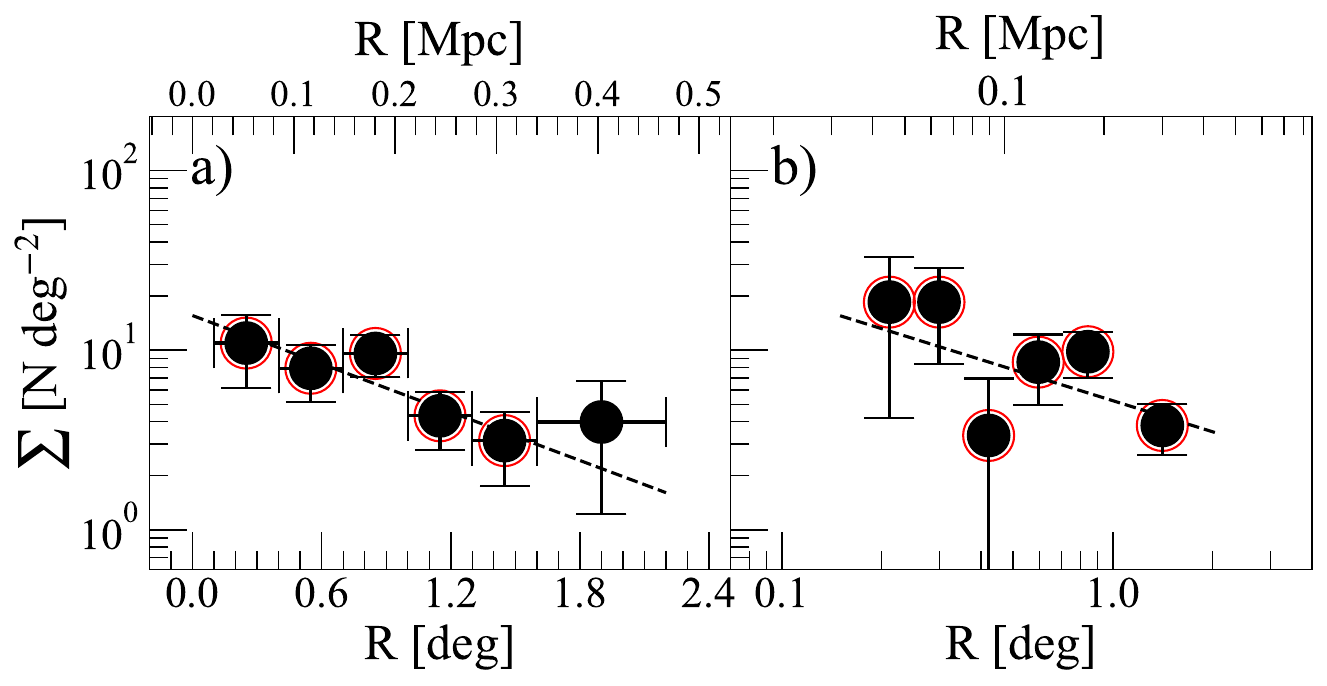}
\caption{Radial number density profiles
of the 55 DGCs in the NGC~2997 group 
in linear (left) and logarithmic (right) scale in x-axis. 
The dashed line in panel a) shows the best-fit exponential function 
for R $\lesssim$ 1.6 \degr\ (see \S\ref{subsec:RND})
obtained by fitting the first five data points, 
excluding the last one.
The dashed line in panel b) shows the best-fit power-law function
obtained using all the six data points in the panel.
\label{fig:RND}}
\end{figure}

Figure~\ref{fig:RND} shows the projected radial number density 
distribution (filled circles) of DGCs in the NGC~2997 group 
decreasing with radial distance after subtracting 
the expected background contamination, 1.75 DGCs deg$^{-2}$ 
(see \S\ref{subsec:sea}).
We adopt both linear and logarithmic radial distance binnings 
in panel a) and b), as was done in previous studies 
\citep{2017Park, 2019Park}.
We fit the exponential function $\Sigma$ = $e^{\alpha \rm R + \beta}$, 
where R is the projected distance, to the number density distribution,
and we obtain the best-fit parameters $\alpha$ = --1.03 $\pm$ 0.40 
and $\beta$ = 2.75 $\pm$ 0.37, shown as the dashed line in panel a).
In the fitting, we exclude the last data point from the outermost bin 
at R = 1.9\degr\ to minimize the effects of the unexpected concentration 
of DGCs in the south-eastern corner of the field 
(see \S\ref{subsec:sea}).
The dashed line in panel b) represents the best-fit power-law function  
$\Sigma$ = ${\alpha \rm R^{\beta}}$,
with $\alpha$ = 5.23 $\pm$ 1.03 and $\beta$ = --0.57 $\pm$ 0.33.

\begin{figure}
\center
\includegraphics[width=0.48\textwidth]{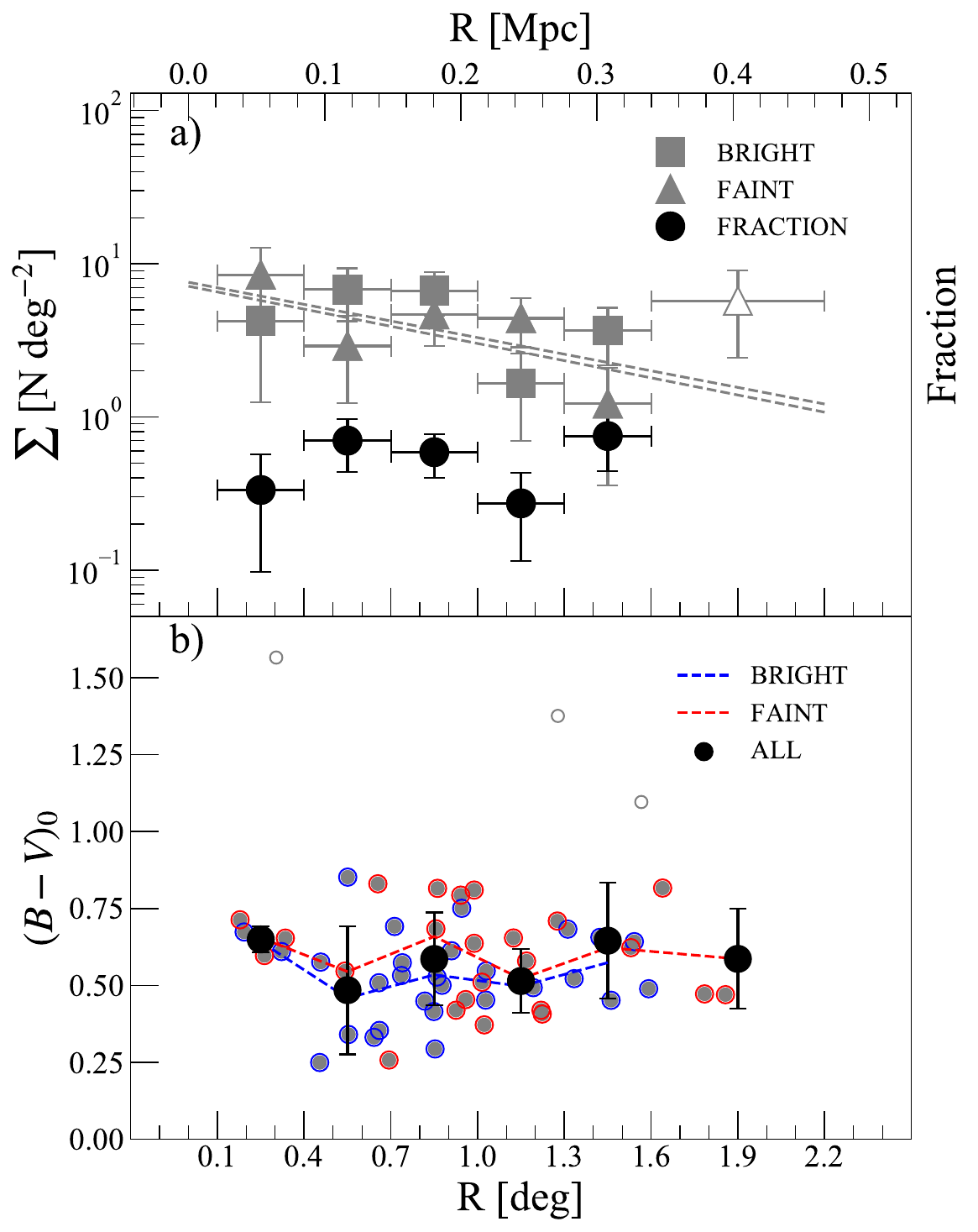}
\caption{{\it Upper Panel}: Radial number density distributions of 
bright (squares) and faint (triangles) groups of the DGCs
in the NGC~2997 group divided by M$_{V}$ = --11.02 mag.
The circles show the distribution of the bright-to-faint ratio.
{\it Lower Panel}: Radial color distribution of the DGCs.
The small circles show the radial distances 
and the \bvo\ colors of the entire 55 DGCs population. 
Those encased by red and blue circles are 
the faint and bright members, respectively. 
(Three out of fifty-five are open circles, while the rest are filled).
The black solid circle, red and blue dashed lines are
the mean color distributions of the entire sample,
faint sample, and bright sample, respectively.
The three small unfilled circles are excluded in
the mean color calculations in order to avoid potential
contributions from outliers.
\label{fig:CRND}}
\end{figure}

Figure~\ref{fig:CRND} compare the radial distributions of the 
binned number density (top panel) and mean \bvo\ color (bottom panel) 
between bright and faint groups of DGCs in the NGC~2997 group 
divided by their median brightness of M$_{V}$ = --11.02 mag.
The radial number density distributions of both groups --- filled squares 
for the bright group and filled triangles for the faint group (top panel),
are largely similar to each other, resulting in a relatively flat 
radial distribution of the fraction of bright galaxies (filled circles).
The best-fit parameters of the exponential function 
($\Sigma$ = $e^{\alpha \rm R + \beta}$)
for the radial number density distribution are 
$\alpha$ = --0.83 $\pm$ 0.54 and $\beta$ = 2.02 $\pm$ 0.51 
(bright group)
and $\alpha$ = --0.86 $\pm$ 0.52 and $\beta$ = 1.96 $\pm$ 0.52 
(faint group).
All the DGCs located at $\rm R$ $>$ 1.6\degr\ belong to the faint group,
and the outermost data point is excluded from the fitting as above.

The binned mean \bvo\ color distributions of both groups --- blue curve 
for the bright group and red curve for the faint 
group (bottom panel) --- largely appear to be constant along the 
radial distance without any significant difference between them, 
resulting in a flat distribution of the mean \bvo\ color for
the 55 DGCs (filled black circles). 
The filled blue and red circles represent colors of all individual DGCs:
blue for the DGCs of the bright group and red for the faint group.
The three DGCs (KSP-DW85, KSP-DW94, and KSP-DW102)
with exceptionally red colors \bvo\ $\gtrsim$ 1.10 mag 
(empty circles; Table~\ref{tab:cat}) are very faint, i.e., 
M$_{V}$ $\gtrsim$ $-$10.35 mag,
and apparently separated from the rest.
They are excluded from the 3-sigma clipping process 
during the calculation of mean colors to minimize the effects of potential outliers.

\subsection{Color and Structure Parameters}\label{subsec:CSP}
\begin{figure}
\center
\includegraphics[width=0.48\textwidth]{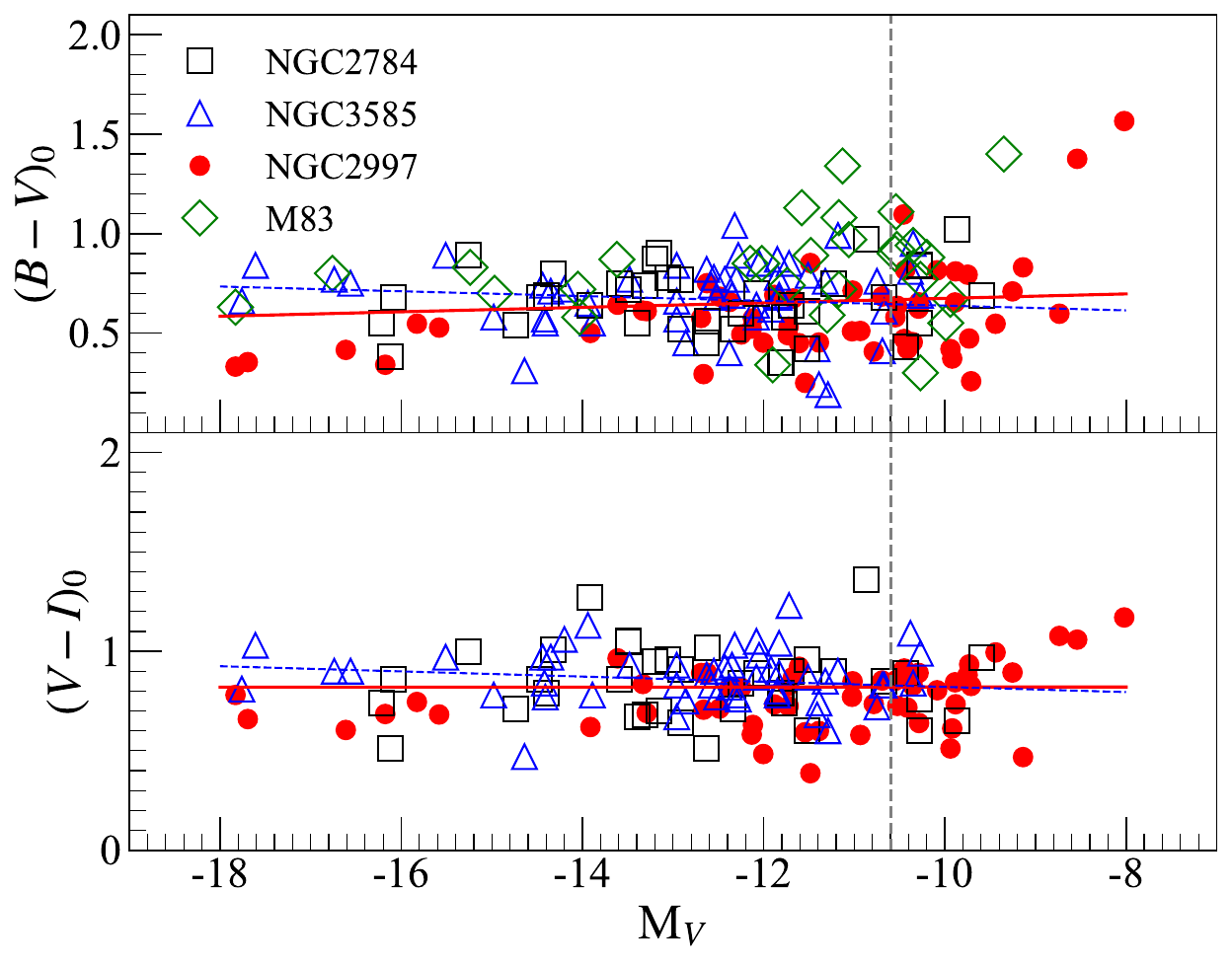}
\caption{Color-magnitude diagrams.
Color-magnitude diagram of the DGCs in NGC~2997 (red filled circles)
compared with those of the NGC~2784 (squares, \citealp{2017Park}), 
NGC~3585 (triangles, \citealp{2019Park}), 
and M83 (diamonds, \citealp{2015Muller}).
The blue dashed line represents the best-fit linear relation 
obtained in \citet{2019Park} 
using the DGCs from both NGC~2784 and NGC~3585 groups.
The red solid line represents the best-fit linear relation
that we obtain from the DGCs of 
NGC~2997, NGC~2784, and NGC~3585 together.
{\tjf The vertical dashed lines indicate the 90\% completeness 
boundaries (see \S~\ref{subsec:sea}).}
\label{fig:CMR}}
\end{figure}

Figure~\ref{fig:CMR} presents the color-magnitude diagrams,
i.e., \bvo\ vs. M$_{V}$ (top panel) 
and \vio\ vs. M$_{V}$ (bottom panel),
for all the DGCs from the four groups:  
NGC~2784 (open squares; \citealp{2017Park}), 
NGC~3585 (open triangles; \citealp{2019Park}),
NGC~2997 (filled red circles), 
and M83 (open diamonds; \citealp{2015Muller}).
The mean values of the \bvo\ and \vio\ colors are respectively 
0.66 $\pm$ 0.16 and 0.84 $\pm$ 0.18 mag for NGC~2784,
0.69 $\pm$ 0.18 and 0.87 $\pm$ 0.14 mag for NGC~3585, and 
0.56 $\pm$ 0.15 and 0.77 $\pm$ 0.16 mag for NGC~2997, 
while the \bvo\ color of M83 is 0.82 $\pm$ 0.25 mag.
The dashed blue lines in Figure~\ref{fig:CMR} represent the
color-magnitude relation (CMR) for the DGCs brighter than M$_{V}$ = --11 mag
in the NGC~2784 and NGC~3585 groups \citep{2019Park}, with slopes of 
--0.012 and --0.013 for \bvo\ and \vio\ colors, respectively. 
These slopes change to $\simeq$ 0 when we also include 
the DGCs from the NGC~2997 group (solid red lines),
indicating that the DGCs in the NGC~2997 group have opposite CMRs
to those of NGC~2784 and NGC~3585.

\begin{figure}
\center
\includegraphics[width=0.48\textwidth]{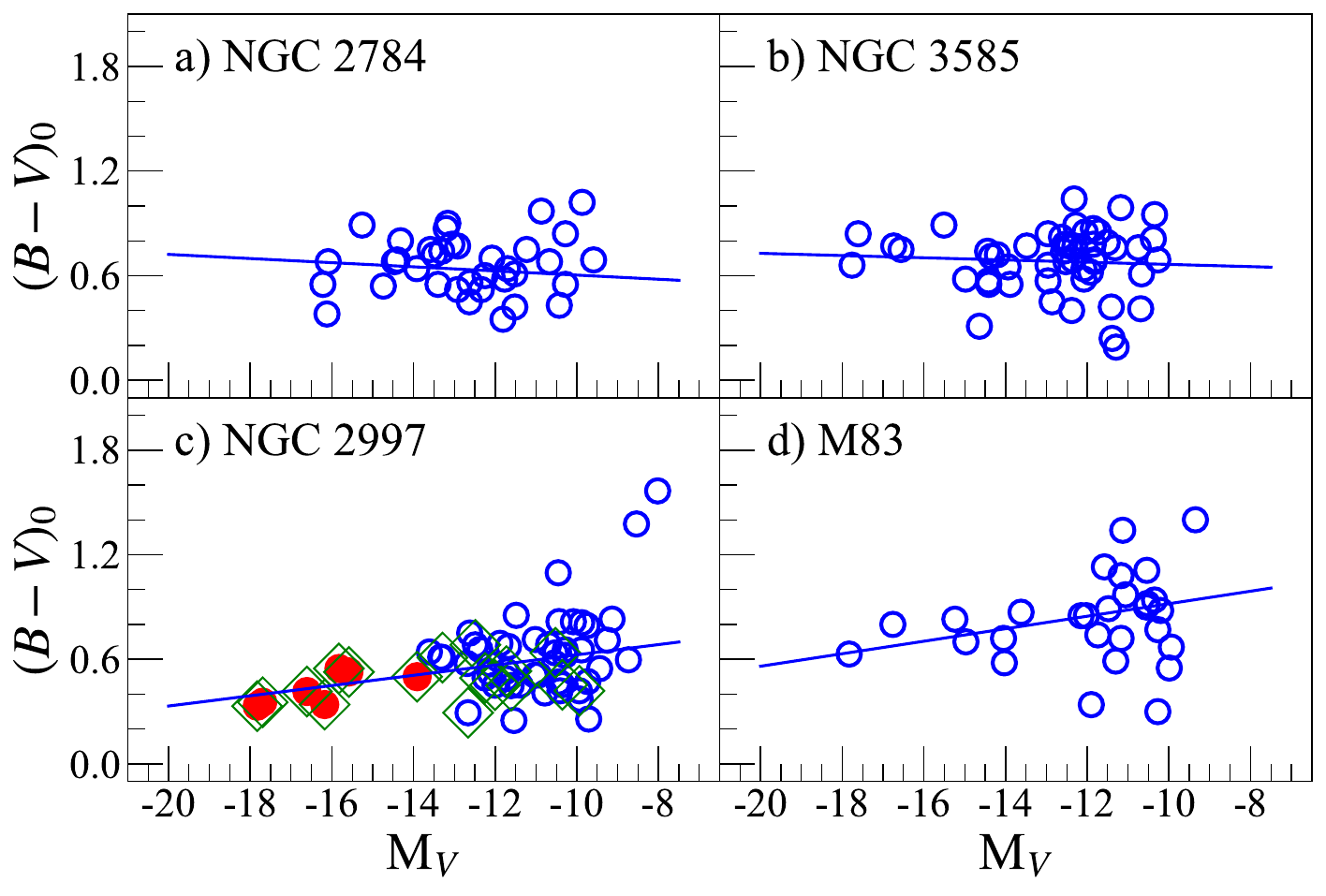}
\caption{\bvo\ vs. M$_{V}$ color-magnitude diagrams of the DGCs in 
NGC~2784 \citep{2017Park}, 
NGC~3585 \citep{2019Park}, 
NGC~2997 (this work),
and M83 \citep{2015Muller} shown in circles.
The blue solid lines represent the best-fit linear relation
with 3-$\sigma$ clipping
that we obtain between the two parameters.
In panel c), the red circles mark the 7 previously
discovered dwarf galaxies (Table~\ref{tab:prev}).
The circles with a green diamond are those 
morphologically identified as irregular, 
while the rest are elliptical.
\label{fig:CMRBV}}
\end{figure}

\begin{figure}
\center
\includegraphics[width=0.48\textwidth]{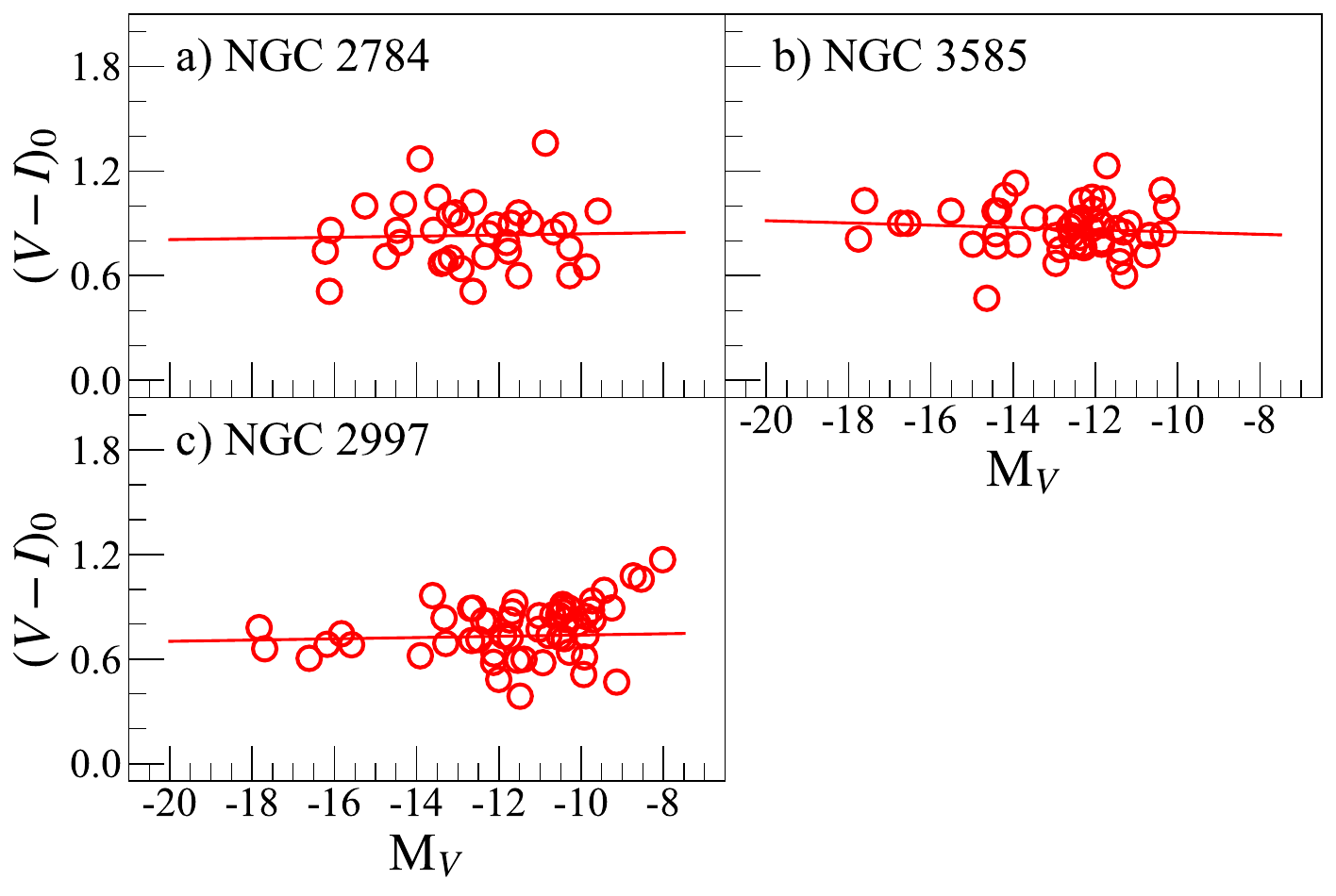}
\caption{Same as Figure~\ref{fig:CMRBV}, but for \vio\ vs. M$_V$
for NGC~2784, NGC~3585, and NGC~2997.
The red solid lines are the best-fit linear relations that we obtain
between the parameters.
\label{fig:CMRVI}}
\end{figure}

We examine the correlation between the two parameters 
(i.e., color vs. absolute magnitude) of the CMR of each group as follows.
For the \bvo\ color (Figure~\ref{fig:CMRBV}), 
the best-fit CMR slopes and the linear correlation coefficients are
--0.01 $\pm$ 0.02 and --0.12 (NGC~2784), --0.01 $\pm$ 0.02 and --0.06 (NGC~3585), 
0.03 $\pm$ 0.01 and 0.41 (NGC~2997), 0.04 $\pm$ 0.02 and 0.33 (M83).
These results show a clear distinction that the brighter DGCs appear bluer
in the two groups with a late-type spiral host (i.e., NGC~2997 and M83), 
while they show an opposite pattern in the other two groups 
with an early-type elliptical host (i.e., NGC~2784 and NGC~3585).
The correlation between the color and magnitude is 
much stronger in NGC~2997 and M83.
For the \vio\ color (Figure~\ref{fig:CMRVI}), 
the best-fit CMR slopes and the linear correlation coefficients are
0.003 $\pm$ 0.02 and 0.03 (NGC~2784), --0.007 $\pm$ 0.01 and --0.08 (NGC~3585), 
0.004 $\pm$ 0.01 and 0.05 (NGC~2997).
We find that correlations in general are much weaker for the \vio\ colors,
and the slopes are essentially flat. 

\begin{figure}
\center
\includegraphics[width=0.48\textwidth]{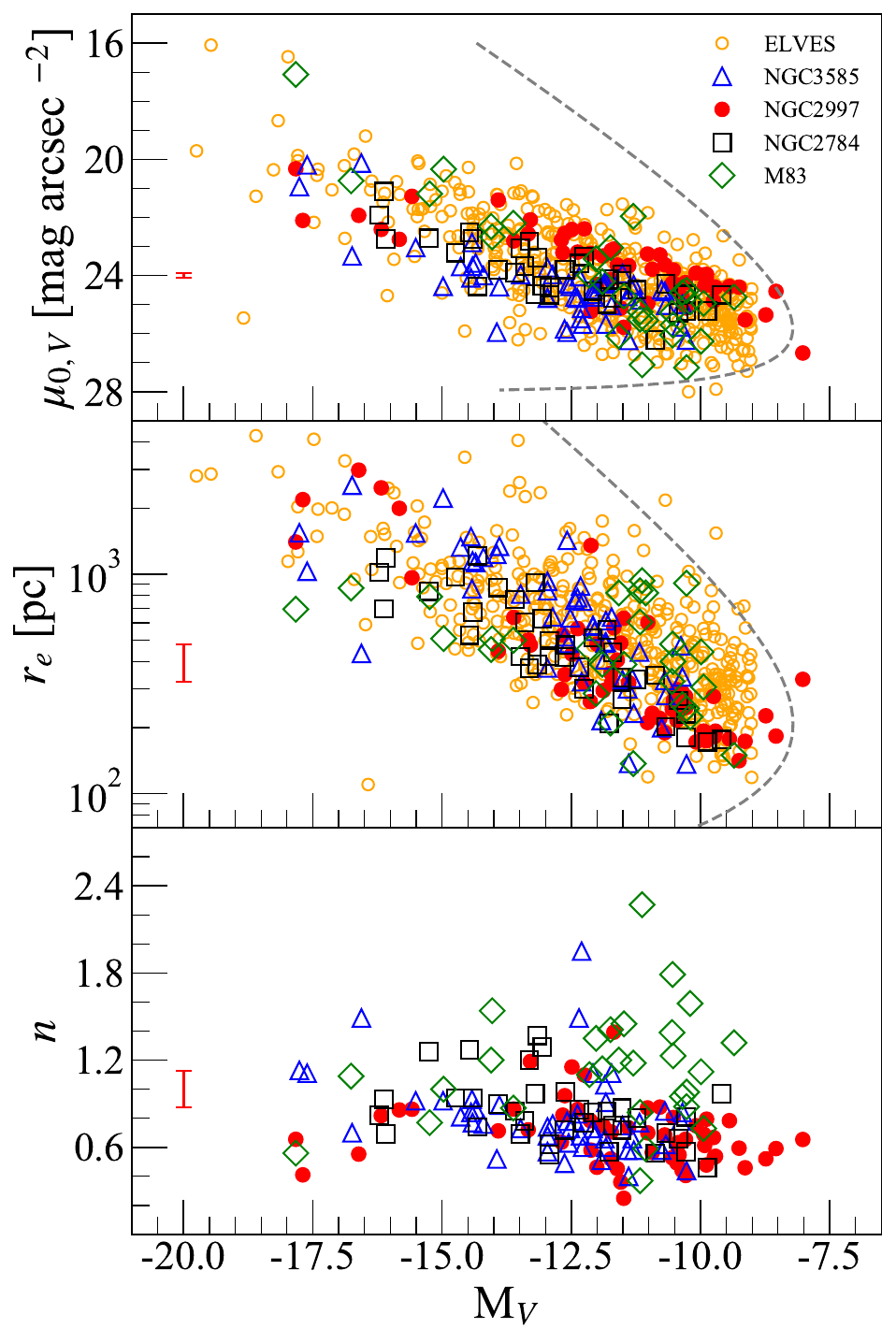}
\caption{Distributions of the apparent central surface brightness
($\mu_{o,V}$; top panel),
effective radii ($r_{e}$; middle panel),
and S\'ersic indices ($n$; bottom panel)
along the $V$-band absolute magnitudes (M$_{V}$)
of the DGCs in NGC~2784 (squares), NGC~3585 (triangles),
NGC~2997 (filled circles), M83 (diamonds).
The gray dashed lines in the top and middle panels are
the completeness boundaries. 
The red error bars in the left-hand part of each diagram
represents the mean uncertainties of the parameters
for the candidates in NGC~2997.
\label{fig:REN}}
\end{figure}

Recent large surveys have estimated structural properties of satellites 
for a large number of MW-like systems 
\citep[e.g. SAGA and ELVES,][]{2021Mao,2021ELVES,2022ELVES}.
In Figure~\ref{fig:REN}, we show the distributions 
of structural parameters---the central surface brightness $\mu_{o,V}$ (top panel),
the effective radius $r_{e}$ (middle panel), 
and the S\'ersic curvature index $n$ (bottom panel)---for DGCs of
NGC~2784, NGC~3585, NGC~2997 from our studies, M83 from \citet{2015Muller}, together with
those of the 444 satellites around MW-like systems from the ELVES survey \citep{2022ELVES}
as a function of their $V$-band absolute magnitudes.
(Note that individual data values of the S\'ersic curvature indices for the ELVES satellites
are unavailable other than a median value of 0.72 \citep{2021ELVES}; 
thus, they are not included in the bottom panel.)
{\tjf As expected, 
we can first identify that most of the DGCs in the 
collected sample reside within the completeness boundary curves 
(gray dashed curves in Figure\ref{fig:REN})
for $\mu_{o,V}$ and $r_{e}$ calculated using the 
following equations from \citet{1988FergusonSandage,2015Muller}:
\begin{equation}
{\rm M}_{V} = \mu_{o,V} - {\rm 5log}(r_{\rm lim}) - 2.18 + 5log(\mu_{\rm lim} - \mu_{o,V})
\end{equation}
\begin{equation}
{\rm M}_{V} = \mu_{\rm lim} - \frac{r_{\rm lim}}{0.5487r_{e, V}} - 2.5log[2\pi\times(0.5958r_{e, V})^{2}]
\end{equation}
where $\mu_{\rm lim}$ and $r_{\rm lim}$ are the limiting surface brightness 
and the limiting radius size for our images and detection.
The equations are based on the assumption that galaxies 
are well-fit by an exponential S\'ersic profile,
which is the case for the 55 DGCs in our study. 
We use $r_{\rm lim}$ $\simeq$ 4\farcs\ for the limiting size, 
which is based on our visual selection criteria (see \S~\ref{subsec:sea}).
We obtain $\mu_{{\rm lim},V}$ = 28 mag arcsec$^{-2}$
for the limiting surface brightness using the following relation from 
\citet{2020Roman} at 3-$\sigma$ level:
\begin{equation}
\mu_{lim} = -2.5 \times {\rm log}(3\sigma/pix\times\Omega)+zero
\end{equation}
where $\sigma$, $pix$, and $\Omega$ represent
the standard deviation of the sky level, 
the pixel size of our images (= 0.4 arcsec, see \S\ref{sec:obs}),
and an angular scale of 8.0 arcsec (2$\times$$r_{\rm lim}$, see \S\ref{subsec:sea})
while $zero$ is the zero point offsets of our images (see Table~\ref{tab:zero}).}
This confirms that the distributions of structure parameters of NGC~2997 DGCs
from our study conform to the predictions on detection limits by \citet{1988FergusonSandage}.
The one DGC located outside the boundary curve is 
DW94 from NGC~2997, and it is very faint (See Figure~\ref{fig:cutoutfull}).
Both the two structure parameters, i.e., $\mu_{o,V}$ and $r_{e}$,
of DGCs from our studies and satellites from ELVES survey
in Figure~\ref{fig:REN} appear to increase as the $V$-band luminosity increases. 
The median value of the S\'ersic curvature index of the DGCs from NGC~2997 
is 0.68, largely consistent with 0.8 for NGC~2784 and NGC~3585, as well as
0.72 for the satellites from the ELVES survey \citep{2021ELVES}. 
It is a bit smaller than 1.1 for M83, but still not significantly different. 
This similarity and consistency of the structure parameters 
from our studies and the ELVES survey for the dwarf galaxies of MW-like systems
is encouraging and may indicate the presence of common patterns in their properties. 
Future investigation of their distributions and properties using 
more statistically significant samples
can lead to advanced understanding of their origin and evolution 
and provide important information for our understanding of 
challenges that the current $\Lambda$CDM model faces \citep[e.g.,][]{2015Weinberg}.

\subsection{Luminosity Function}\label{subsec:LF}
\begin{figure}
\center
\includegraphics[width=0.48\textwidth]{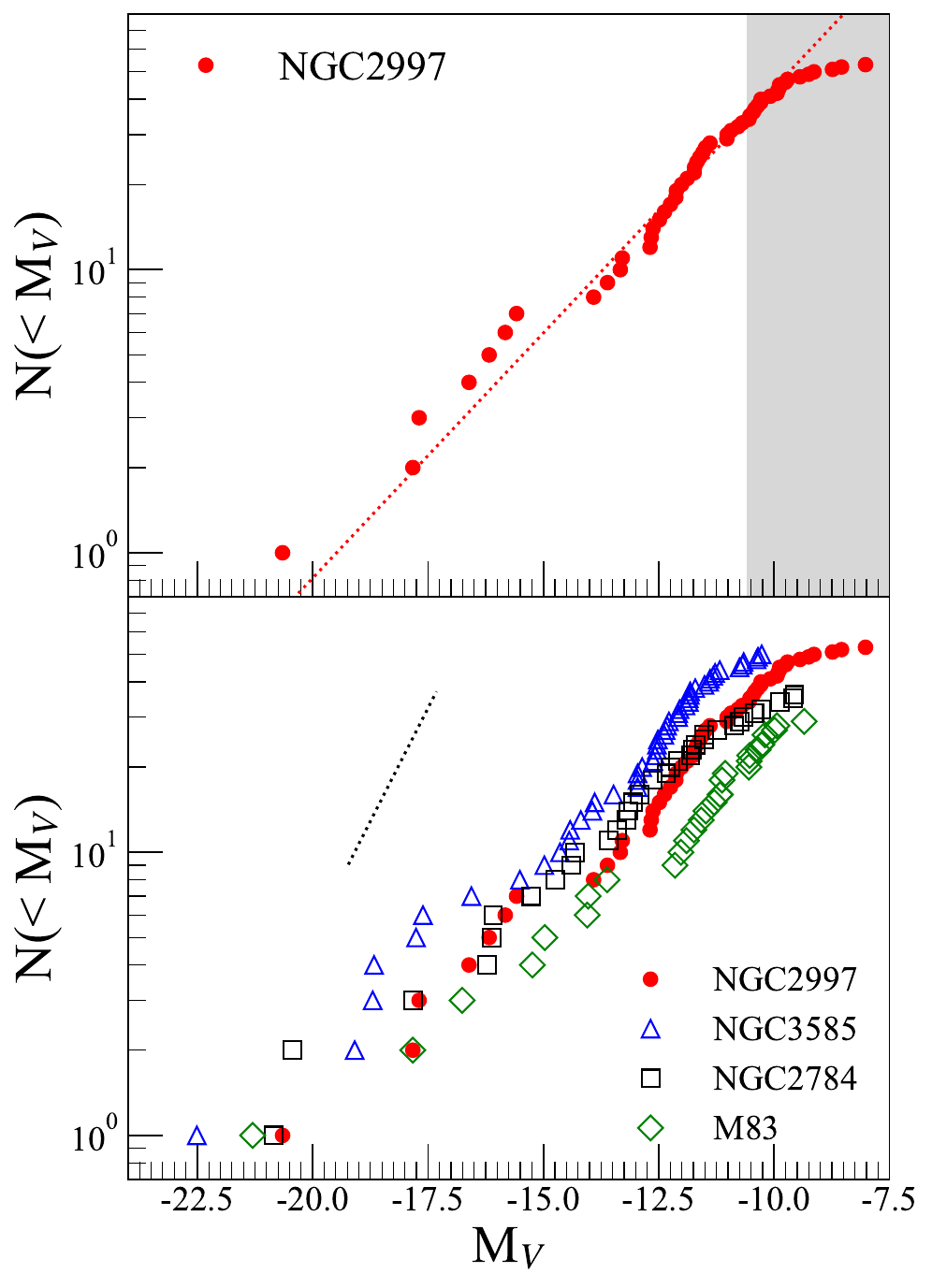}
\caption{({\it Top}) The cumulative LF (red filled circles) 
of the NGC~2997 group. 
The dotted line represents the best-fit Schechter function with 
the faint-end slope $\alpha$ $\simeq$ --1.43.
The shaded area indicates the magnitude region below our work's 90\% completeness magnitude limit 
($M_{V}$ $\simgt$ --10.6 mag).
({\it Bottom}) The cumulative LFs of NGC~2997 compared with those of
NGC~2784 \citep[squares;][]{2017Park}, 
NGC~3585 \citep[triangles;][]{2019Park},
and M83 \citep[diamonds;][]{2015Muller}.
The black dotted line is the case of $\alpha$ = --1.8
predicted by the $\Lambda$CDM model \citep{2002Trentham}. 
\label{fig:LF}}
\end{figure}

In Figure~\ref{fig:LF} (top panel), we show
a cumulative LF of the NGC~2997 group
constructed with 52 DGCs that we used for the radial number
density distribution analysis (Figure~\ref{fig:RND}),
excluding three DGCs in the last radial bin,
together with M$_{V}$ $\simeq$ --20.65 mag 
for NGC~2997 itself \citep{1988dVau}.
The 52 DGCs are at projected distances $\lesssim$ 400 kpc, 
which is the virial radius of the group, 
and include all of the previously-discovered members.
The dashed line in the top panel of Figure~\ref{fig:LF} 
represents the best-fit Schechter function 
with a faint-end slope of $\alpha$ = --1.43 $\pm$ 0.02
obtained by fitting the LF where M$_{V}$ $\leq$ --10 mag, 
as was done in our previous studies, 
using the non-linear least-square fitting method.
The shaded region shows the area for M$_{V}$ $\simgt$ --10.6 mag, which
is the 90\% completeness magnitude limit of our detection (\S~\ref{subsec:sea}).
The slope of the LF for the faint dwarf galaxy candidates 
in this area becomes more flattened as the luminosity decreases, 
indicating that the flattening is at least partially attributed by the
incompleteness of our detection for $M_{V}$ $\simgt$ --10.6 mag.

{\tjf We explore the effects of background contamination, 
which become more substantial for the population of faint dwarf galaxies, 
to our measurement of the faint-end LF slope above.
According to the recent results from the ELVES survey,
low luminosity (M$_{V}$) dwarf galaxies with higher central surface brightness 
($\mu_{0,V}$) are more likely to be contaminants than those
with lower central surface brightness \citep{2022ELVES}.
As described in \S~\ref{subsec:sea}, 
we expect 7--16 background contaminants out of the 55 DGCs
that we identified in the NGC~2997 field.
If we assume that there are 16 contaminants in our sample of 55 DGCs
and that all of them follow the same pattern of stellar-mass
dependent contamination-likelihood found by the ELVES survey,
we estimate the contamination-corrected LF slope 
of the NGC~2997 group to be --1.23 $\pm$ 0.03.
For this, we identify 16 DGCs in our sample that are more likely 
contaminants than others and exclude them in the LF slope fitting as follows:
we first establish a linear relation between M$_{V}$ and $\mu_{0,V}$ among the 55 DGCs,
and then select 16 DGCs that have largest excess
in $\mu_{0,V}$ to the linear relation. 
We  only select DGCs that are fainter than M$_{V}$ = --12.5 mag, 
which is the expected upper magnitude limit for our background contaminants  
(see \S~\ref{subsec:sea}). 
The final 16 DGCs chosen as contaminants 
have excess $\mu_{0,V}$ by more than 0.25 mag arcsec$^{-2}$
when compared to the expected values from the linear relation.
The background contamination-corrected faint-end slope is
about 0.2 shallower than the slope measured without such correction. 
Since 16 is the upper limit for the number of background contaminants,
we expect the true value of the faint-end LF slope after 
stellar-mass dependent correction to be between --1.43 and --1.23.}

The bottom panel of Figure~\ref{fig:LF} compares 
the cumulative LF of NGC~2997 with those of 
NGC~2784 \citep{2017Park}, 
NGC~3585 \citep{2019Park},
and M83 \citep{2015Muller}.
Applying the same fitting method, 
we obtain their best-fit LF faint-end slopes to be
--1.22 $\pm$ 0.03 (NGC~2784),
--1.33 $\pm$ 0.03 (NGC~3585), 
and --1.40 $\pm$ 0.04 (M83).
The faint-end slope of NGC~2997 is very similar to that of M83, 
while it is a bit steeper than those of NGC~2784 and NGC~3585. 
Compared to the range of --1.9 $\lesssim$ $\alpha$ $\lesssim$ --1.6 
typically predicted by $\Lambda$CDM models 
\citep[e.g.,][]{1999Klypin, 2002Trentham, 2008Springel, 2016Han, 2019PengfeiLi},
the faint-end LF slopes of these galaxies appear to be consistently flatter.
However, we note that recent $\Lambda$CDM simulations 
based on different types of stellar-to-halo mass relation,
which is still poorly constrained for low-mass dwarf galaxies, 
have shown that there exists a great uncertainty in the predicted LFs \citep[e.g.,][]{2021Carlsten}.
It is, therefore, premature to ascertain whether the slope that we measure 
for the NGC~2997 group is in an agreement or disagreement with $\Lambda$CDM model prediction.

\begin{figure}
\center
\includegraphics[width=0.48\textwidth]{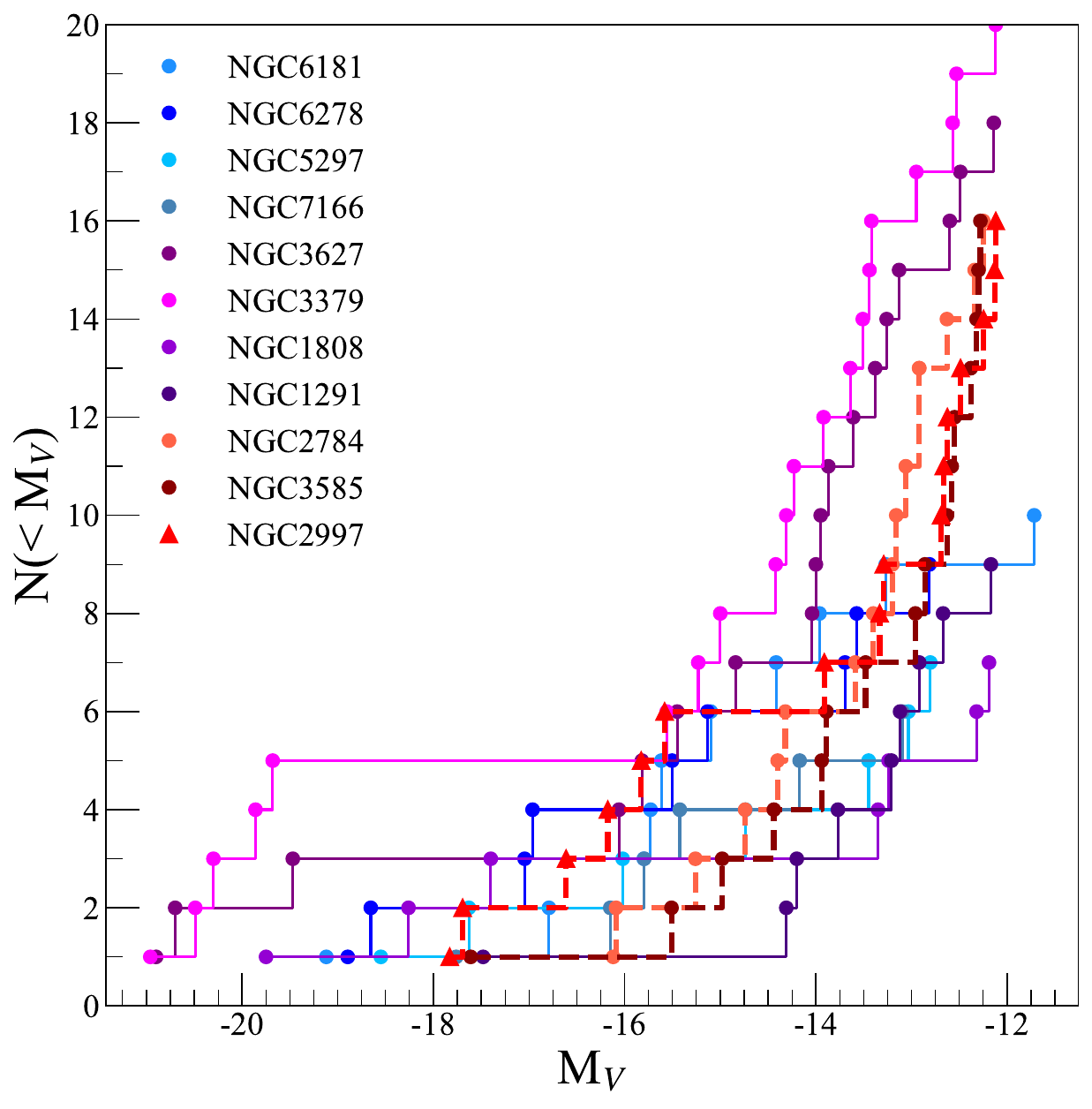}
\caption{Luminosity function of NGC~2997 compared with those of sample 
galaxies from the SAGA and ELVES surveys as well as from our previous studies (see \S~\ref{subsec:LF}).
The purple-hued points and lines are the groups from the ELVES survey, while the 
blue-hued points and lines are the groups from the SAGA survey. 
{\tjf The red points and dashed lines are the groups from the KSP survey. }
\label{fig:MetaLF}}
\end{figure}

Figure~\ref{fig:MetaLF} shows the distributions of the LFs
for bright ($M_{V}$ $\simlt$ --12.1 mag) satellites 
around 11 MW-like galaxies: 
four (NGC~6181, 6278, 5297, and 7166; shown in blue-hued circles and lines) from the SAGA survey, 
four (NGC~3627, 3379, 1808, and 1291; purple-hued circles and lines) from the ELVES survey, and 
three (NGC~2784, 3585, and 2997; red-hued symbols and dashed lines) from the KSP (including this study). 
The SAGA groups consist of the 4 most populated groups from the survey
in terms of the number of confirmed satellites,
while the 4 groups from the ELVES survey are selected because they have
distances similar to that of NGC~2997 with 
a satellite survey size comparable to that of the SAGA survey, i.e., R $\simlt$ 300 kpc.
We can identify in the figure that the LFs of the three KSP group members,
which are similar to each other,
lie between those of the ELVES groups for faint regime of M$_{V}$ $\simgt$ --14 mag
with more rapid increase in number of satellites as they become fainter.
The LF slopes of the KSP and the two ELVES groups appear to be much steeper than
that of the SAGA survey in this faint regime. 
According to \citep{2022ELVES}, this might have been caused by incompleteness 
beyond what is accounted for in the SAGA survey, 
especially in the low-mass regime (see the reference for the details).
As we explained in \S~\ref{subsec:sea},
the detection rate of bright satellites from our work on NGC~2997 is
similar to that of the SAGA survey, which we can confirm in their LFs in Figure~\ref{fig:MetaLF}.
Two galaxy groups, which are NGC~3379 and NGC~3627, from the ELVES survey 
have brightest ($<$ --20 mag) satellites as well as the largest number of faint ones,
confirming that groups with brighter satellites tend to have a larger number of 
satellites, as have been found in recent studies \citep{2021Mao,2022ELVES}.

{\tjf There exists no DGC in our sample that has M$_{V}$ around --15 mag (Table~\ref{tab:cat}),
which we identify as a 2-mag gap-like feature near 
M$_{V}$ = --15 mag in the LF.}
In the SAGA survey of small galaxy groups \citep{2017Geha},
the majority of the groups show similar magnitude gaps
in the range of 2.2--3.6 mag,
consistent with the prediction by numerical simulations
presented in the same work, while a small number of faint 
(i.e., M$_{K}$ $\simgt$ --23.75 mag for the host galaxy luminosity)
groups have gaps greater than 5 mag.
{\tjf Considering that NGC~2997 is MW-analogous with M$_{K}$ $\simeq$ --24.02 mag 
(\S\ref{sec:intro}) similar to the systems studied in the SAGA survey, 
the 2.2 mag gap-like feature in the LF of NGC~2997 may indicate 
the presence of a similar gap among small MW-analog groups \citep[see also][]{2019Bennet}. 
In order to examine the statistical significance of the gap-like feature in NGC~2997,
we perform a p-value test on 100,000 cases of randomly-sampled 55 dwarf galaxies 
from a Schechter function constructed by using the LF parameters of NGC~2997 group.
As a result, we find that about 60~\% of the 100,000 cases have the presence 
of such a gap in their LFs, meaning the gap-like feature in the LF of NGC~2997
is more likely a result of random sampling with small statistical significance.}

\section{Evolutionary stage of the NGC~2997 group}\label{sec:DE}

The DGCs that we identify in the field of NGC~2997 show several 
notable characteristics diagnostic of their evolutionary stage
as we discuss below.

\subsection{Color-Magnitude Relation, Star Formation Activities, 
and Galactic Conformity}\label{subsec:cmr}
The CMR of the DGCs in NGC~2997 indicates that 
the group is populationally young with ongoing star formation activities.
Their \bvo\ colors show a positive correlation with 
M$_{V}$ (Figure~\ref{fig:CMRBV}; \S\ref{subsec:CSP})
with brighter members being bluer,
and this suggests the presence of younger stellar populations 
in the brighter and more massive members,
indicating that the star-formation activities in less 
massive members are more quenched than their massive counterparts.
Since most of our candidates are within the virial radius ($\simeq$ 400kpc) of NGC~2997, 
it is likely that their quenching is caused by the group environmental effects.
This interpretation is further supported by various observational results
showing a clear contrast in quenching between field and satellite dwarf galaxies:
the former have shown an extreme lack of quenching \citep[e.g.][]{2011Weisz,2012Geha}, 
whereas the latter have shown much stronger environmental quenching 
\citep[e.g., see][and references therein]{2018Fillingham}.
Given that the timescale of environmental quenching is much shorter 
for lower mass satellites \citep[e.g.,][]{2022Samuel},
it is conceivable that only the low-mass members of the group have 
had enough time for the environmental quenching to develop, 
causing them to appear redder than the massive members.

It is known that colors of shorter wavebands (e.g., \bv) 
are more affected by star formation activities 
(or stellar age variations) than those in longer wavebands 
\citep[e.g., \vi, see][for example]{2006Chang, 2009Janzetal}.
The \vio\ colors of DGCs in the NGC~2997 group
show a much weaker correlation with their {\tjf luminosity}
than \bvo\ color does (\S\ref{subsec:CSP}), 
supporting the interpretation that the CMR of NGC~2997 
could be primarily driven by star formation activities, 
and hence stellar age variations. 
Similar relations between the color and magnitudes of satellite galaxies 
to those of NGC~2997 have been found in many recent observational and simulation studies.
More massive satellites found in the ELVES survey around MW-like hosts
appear systematically bluer than the less massive members \citep{2022ELVES},
consistent with the measurements made by H$\alpha$
showing that the quenched fraction decreases as the mass of satellites increases.
Recent FIRE-II simulation of cosmological galaxy formation \citep[e.g.][]{2022Samuel}
has also found that massive satellites around MW-like galaxies 
have significantly lower quenched fractions compared to less massive members.
The results from these observational and simulation studies are in line with 
our interpretation that the more massive DGCs of NGC2997 
have experienced less quenching.
Additionally, there also exists a higher fraction of irregular morphologies 
among the brighter DGCs in NGC~2997 
(\S\ref{subsec:NUG} and \S\ref{subsec:mor}),
further supporting this interpretation since irregular galaxies in general
appear to have more persistent star formation activities
\citep{2009Chiboucas,2012Kormendy,2013Kirby,2021Lassen,2021Scott}.
These observed properties of the DGCs in NGC~2997 suggest that 
the brighter dwarf galaxies in NGC~2997 are more dominated by 
star-forming late-type galaxies with younger stellar age.

The average $(g-r)_{0}$ color of the 127 satellite galaxies 
identified in the SAGA survey around 36 MW-analog hosts  
with the detection limit of M$_{r}$ $\simlt$ --12.3 mag 
is 0.39 $\pm$ 0.13 mag \citep{2021Mao},
which converts to \bv\ $\simeq$ 0.61 $\pm$ 0.13 mag when adopting
the conversion relation between $g-r$ and \bv\ colors \citep{2005Jester}.
(Note that, according to \citep{2005Jester},
the rms conversion uncertainty between these two colors is small at the level of 0.04 mag.)
In our study, the average \bvo\ color of the 18 brightest DGCs of NGC~2997
with M$_{V}$ $\simlt$ --12 mag, which is equivalent to the SAGA detection limit above, 
is 0.52 $\pm$ 0.13 mag.
The bright DGCs in NGC~2997 have comparable \bvo\ colors
to those of the SAGA satellites.
Considering that most of the SAGA satellites 
have apparent star formation activities, 
this could also indicate that the bright DGCs in NGC~2997 have 
active star formation activities,
consistent with our interpretation that 
they are mostly star-forming late-type galaxies.

The CMR pattern of NGC~2997 from our study appears to be
different from those of NGC~2784 and NGC~3585 
from our previous studies 
in that the \bvo\ colors and luminosity of the DGCs 
of the two early-type hosted groups
show weak, negative correlations \citep{2017Park, 2019Park}.
We note, however, that the DGCs of the M83 group, 
which has a late-type host like NGC~2997,
show a strong, positive correlation between \bvo\ and {\tjf luminosity}, 
similar to NGC~2997 (Figure~\ref{fig:CMRBV}), 
indicating more active star-formation activities 
in more massive members.
It appears, therefore, that the DGCs 
in the groups of NGC~2784, NGC~3585, and NGC~2997 as well as M83
show star-formation activities 
that are overall similar to those of their hosts,
compatible with ``galactic conformity'' wherein 
satellite galaxies tend to show similar 
star-formation activities and colors 
to their hosts \citep{2006Weinmann, 2016Bray, 2018Treyer}.

\subsection{Radial Distribution of Mass and Color}\label{subsec:radial}
{\tjf As seen in Figure~\ref{fig:CRND},
there appears to be no apparent segregation in mass and color 
of the DGCs in NGC~2997, showing that the DGCs near the center of the group 
are neither particularly more massive nor redder.
In some cluster and groups, the central area near the host is 
predominantly populated by more massive satellites 
\citep{2012Presotto,2015Roberts,2016Joshi,2018Barsanti,2020KimSw}, 
often attributed to the stronger effects of dynamical friction \citep{1943Chandrasekhar}. 
However, some previous studies have also found no signs of mass segregation 
in other clusters and groups \citep{2010vonderL,2013Vulcani,2016Kafle}. 
It has been suggested that gas-stripping environmental effects,
i.e. ram pressure and tidal stripping \citep{2016Kafle,2018Barsanti,2020KimSw},
could lead to less efficient dynamical friction and mass segregation. 
In the case of NGC~2997 group,
which is located in a low-density environment of 
the loose group LGG 180 (see \S~\ref{sec:intro}) 
without any detected diffuse X-ray emission \citep{2006Sengupta},
the gas stripping processes are likely relatively inefficient, 
which is compatible with the known insignificant deficiency of \hi\ gas
in LGG~180 galaxies \citep{2006Sengupta,2012Pisano,2014Pisano},
making dynamical friction more prevalent.
We estimate the dynamical friction time-scale 
for brighter (M$_{V}$ $\simlt$ --15 mag) and fainter members
of the NGC~2997 group to be about 7 Gyrs in average 
and longer than the Hubble time, respectively,
assuming a uniform light-to-mass ratio and following the method in \citet{2008Jiang}. 
One feasible scenario explaining the absence of mass segregation 
in the NGC~2997 group is that its massive members
have been accreted onto the group halo recently during the last 7 Gyrs. 
Similar to our scenario, some studies \citep[see][for example]{2018Barsanti} 
of small group environments (like NGC~2997, \S~\ref{sec:intro})
have strongly suggested that massive star-forming satellites 
are likely new members recently accreted from surrounding fields. 
Since recently accreted field dwarf galaxies have 
higher star-formation activities and bluer colors \citep[e.g.][]{2011Weisz,2012Geha,2018Barsanti}, 
this interpretation is also consistent with our CMR (see \S\ref{subsec:cmr}).}

The \bvo\ color distribution of the DGCs in NGC~2997 
along the radial distance from the host is overall flat
(Figure~\ref{fig:CRND}, bottom panel)
without any apparent radial dependency,
which is in contrast to what can be expected for an evolved system 
where earlier and stronger quenching of star formation 
driven by stronger tidal forces and also by more effective interactions 
with the intracluster medium \citep{2013Wetzel,2017Park}
would cause satellite galaxies near the center to be redder 
than those at larger distances \citep[e.g.,][]{2012Presotto,2013Lisker,2016LEE}.
For example, according to the study by \citet{2012DeLucia} based on the Millennium Simulation,
a much steeper radial color distribution is plausible 
for dwarf galaxies that have been satellites for over at least 5--8 Gyrs.
{\tjf The absence of color segregation among the NGC 2997 DGCs, therefore,
could be consistent with the scenario that some of them are newly accreted members. 
This is also inline with our interpretation above for the absence of
mass segregation in which we suggest that the massive members of NGC~29997,
which are blue (Figure~\ref{fig:CMRBV}), are recently accreted.}
Another possible explanation for the origin of the observed shallow color gradient
of NGC~2997 is less efficient quenching for a late-type system 
\citep{2006Weinmann, 2016Bray, 2018Treyer} in that 
a shallow color gradient may simply results from similar star 
formation activities between the host and satellites, known as ``galactic conformity." 
{\tjf However, it is not clear how this process can explain 
the observed absence of mass segregation in NGC~2997.
Overall, the absence of mass and color segregation in NGC~2997
appears to be compatible with the scenario that the system 
has recently accreted massive, blue dwarf galaxies, 
possibly from the surrounding field.}

\subsection{Morphologies}\label{subsec:mor}
Another notable characteristic of the DGCs of NGC~2997 group is their morphology.
While we identify only one UDG and one nucleated dwarf galaxies
(see \S\ref{subsec:NUG}) in the NGC~2997 group, 
ten and four of the DGCs in NGC~3585 are nucleated and UDGs, 
respectively \citep{2019Park}.
In addition, as shown in \S\ref{subsec:sea} and Figure~\ref{fig:CMRBV},
NGC~2997 appears to have a higher ratio of morphologically irregular 
dwarf galaxies (dIs) --- the fraction of dIs is 31\% $\pm$ 9\% in NGC~2997, 
while it is 18\% $\pm$ 8\% and 11\% $\pm$ 7\% in NGC~2784 and NGC~3585 when adopting 
the same selection criteria (See \S~\ref{subsec:NUG}), respectively. 
These potential differences in the morphology of the DGCs 
between the NGC~2997 system and the other two appear to be consistent 
with our interpretation that this system is in a low-density environment, 
and it is populationally young with recent accretion of new dwarf galaxies.
First, the relative paucity of the nucleated dwarf galaxies 
can be understood in populationally young satellite galaxies 
where there has not been enough time for globular clusters 
to infall to form nucleus \citep{2000OhLin,2021Fahrion,2021Poulain}. 
Secondly, it might be difficult to develop UDGs 
in a low density environment (LGG~180) via tidal stripping of gas
and/or their size expansion through tidal heating \citep{2020Tremmel,2021Jones}. 
Thirdly, the higher fraction of dIrrs among the DGCs in NGC~2997 
is compatible with a them being newly accreted dwarf galaxies where there has not 
been enough time for them to evolve into more early-type morphologies 
\citep{2003Pasetto,2010Dunn,2015Chattopadhyay,2021Scott}.

\section{Summary and Conclusion}\label{sec:sum}
In the work presented in this paper, we conduct photometric analysis
of 55 DGCs that we have identified 
in the field of the giant spiral galaxy NGC~2997 
using the deep stacked images 
obtained from the KMTNet Supernova Program.
Out of 55, 48 are newly discovered in this work 
while the rest seven are already identified candidates from previous studies.
Table~\ref{tab:comp} compares the parameters of the DGCs 
in the NGC~2997 group that we identify in this study 
to those of the NGC~2784 and NGC~3585 groups.
We summarize our main results below, along with our conclusions
about the evolutionary stage of the NGC~2997 group based on our results.

\begin{itemize}

\item
The 55 candidates show a broad range of 
photometric and structural parameters 
that are overall compatible with those found 
from the dwarf galaxies in other galaxy groups.
Their $I$-band effective radii, $I$-band S\'ersic index,
and the $V$-band total absolute magnitudes range from 0.142 to 2.97 kpc,
from 0.25 to 1.39, and from --8.02 to --17.69 mag, respectively.
The faint-end slope of their luminosity function is $\alpha$ = --1.43 $\pm$ 0.02,
which is shallower than predictions based on $\Lambda$CDM models,
but largely comparable with previously measured values. 
{\tjf (\S~\ref{subsec:SBPNCAT},\S~\ref{subsec:LF})}

\item 
The DGCs of the NGC~2997 group are bluer than those 
of the NGC~2784 and NGC~3585, 
with its brighter members being systematically bluer 
than the fainter ones.
The inferred higher star-formation activities in the 
more massive members of the NGC~2997 group
could result from {\tjf them having experienced 
group environmental quenching for a shorter time. (\S~\ref{subsec:cmr})}

\item 
The projected number density of the DGCs in NGC~2997 group decreases 
along the distance from the center of NGC~2997,
and there is no difference in this radial number density distribution
between the bright (M$_{V}$ $\lesssim$ --11.02 mag) 
and faint members of the group.
The average color of them appears to be invariant 
as a function of the radial distance. 
{\tjf The lack of mass or color segregation again appears to be compatible with 
the scenario that the system has recently accreted massive, blue dwarf galaxies, 
possibly from the surrounding field. (\S~\ref{subsec:radial})}

\item
Compared to our previous studies of DGCs 
in the field of NGC~2784 and NGC~3585
using similar KMTNet data, the DGCs of NGC~2997 
show a lack of more morphologically developed candidates --- out of 
the 55 candidates, we are able to identify only 
one ultra-diffuse dwarf galaxy, and one nucleated candidates.
In contrast, there exists a significantly increased fraction 
of irregularly shaped candidates (17 in total) in our sample of NGC~2997, 
and they are significantly more luminous and bluer. 
These morphological distributions of the DGCs indicate
that the NGC~2997 group is populationally young {\tjf with
recently accreted luminous dwarf galaxies. (\S~\ref{subsec:mor})} 

\item 
For the brighter DGCs in NGC~2997, we find their detection rate
and the inferred star formation activities are largely comparable 
to those found in the SAGA survey,
whose detection limit is M$_{V}$ = --12.1 mag,
of MW-analog groups in the distance range 25--42 Mpc. 
Similar to the results from the SAGA and also to the ELVES survey, 
the faint DGCs in NGC~2997 appear to be 
substantially more quenched than bright ones. 
Identification of whether this trend of quenched star formation activities
among faint DGCs in NGC~2997 is a general property of 
faint satellites around MW-analog hosts requires more samples. 

\end{itemize}

\section*{Acknowledgments}

We are very grateful to the anonymous referee for the thorough and 
valuable comments that helped improve this paper significantly.
This research has made use of the KMTNet system operated by the 
Korea Astronomy and Space Science Institute (KASI)
at three host sites of CTIO in Chile, SAAO in South Africa, and SSO in Australia.
Data transfer from the host site to KASI was supported by the 
Korea Research Environment Open NETwork (KREONET).
We acknowledge with thanks the variable star observations
from the AAVSO International Database contributed by
observers worldwide and used in this research. 
DSM was supported in part by a Leading Edge Fund 
from the Canadian Foundation 
for Innovation (project No. 30951) and a Discovery
Grant (RGPIN-2019-06524) from the Natural Sciences and 
Engineering Research Council (NSERC) of Canada. 
HSP was supported in part by the National Research Foundation of Korea 
(NRF) grant funded by the Korea government 
(MSIT, Ministry of Science and ICT; No. NRF-2019R1F1A1058228).
Y.D.L. acknowledges support from Basic Science Research Program 
through the National Research Foundation of Korea (NRF) 
funded by the Ministry of Education (2022R1I1A1A01054555).
This research was supported by the Korea Astronomy 
and Space Science Institute under the R\&D program (Project No. 2023-1-868-03)
supervised by the Ministry of Science and ICT.

\section*{Data Availability}
The data underlying this article will be publicly available online
once it is accepted for publication.



\bibliographystyle{mnras}
\bibliography{KSP-DG-2997} 



\appendix
\section{Large Figures and Tables}
\begin{figure*}
\center
{\includegraphics[width = \textwidth]{Figures/CUTOUT1.pdf}}\\
{\includegraphics[width = \textwidth]{Figures/CUTOUT2.pdf}}\\
{\includegraphics[width = \textwidth]{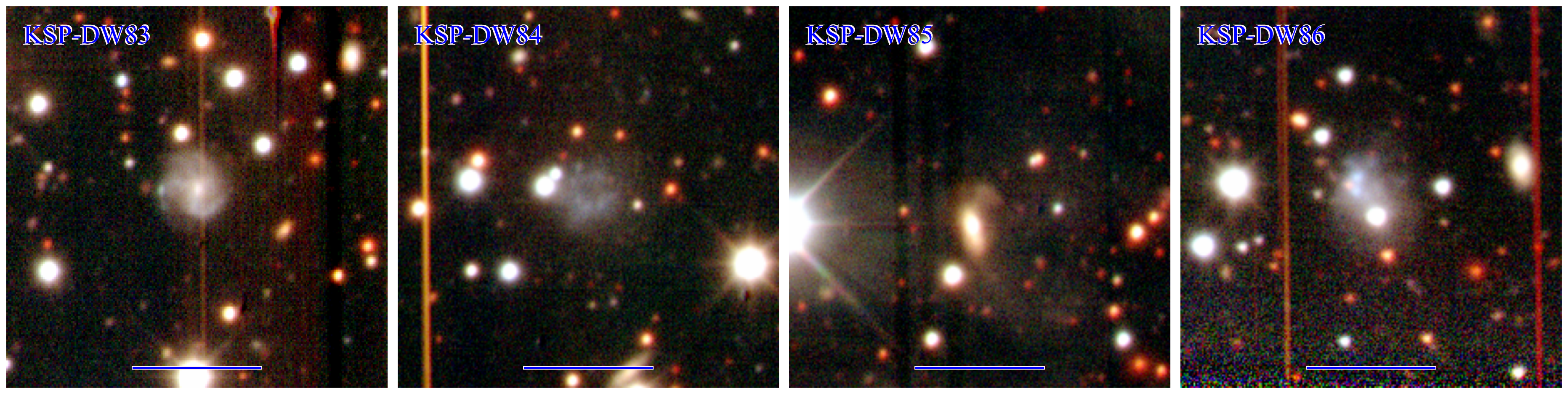}}\\
{\includegraphics[width = \textwidth]{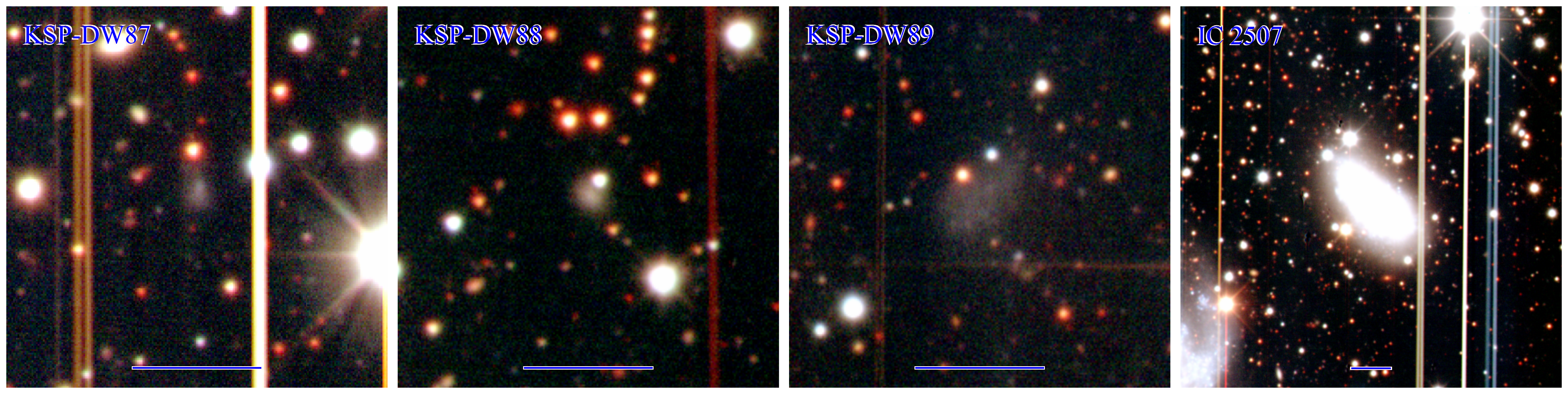}}\\
\caption{RGB cutout images of the 55 DGCs in Table~\ref{tab:cat}. 
North is up and east is to the left. 
The blue horizontal bar at the bottom of each image
corresponds to an angular size of 0\farcm5 
(= 1.77 kpc at the distance of 12.2 Mpc).
\label{fig:cutoutfull}}
\end{figure*}

\begin{figure*}
\center
{\includegraphics[width = \textwidth]{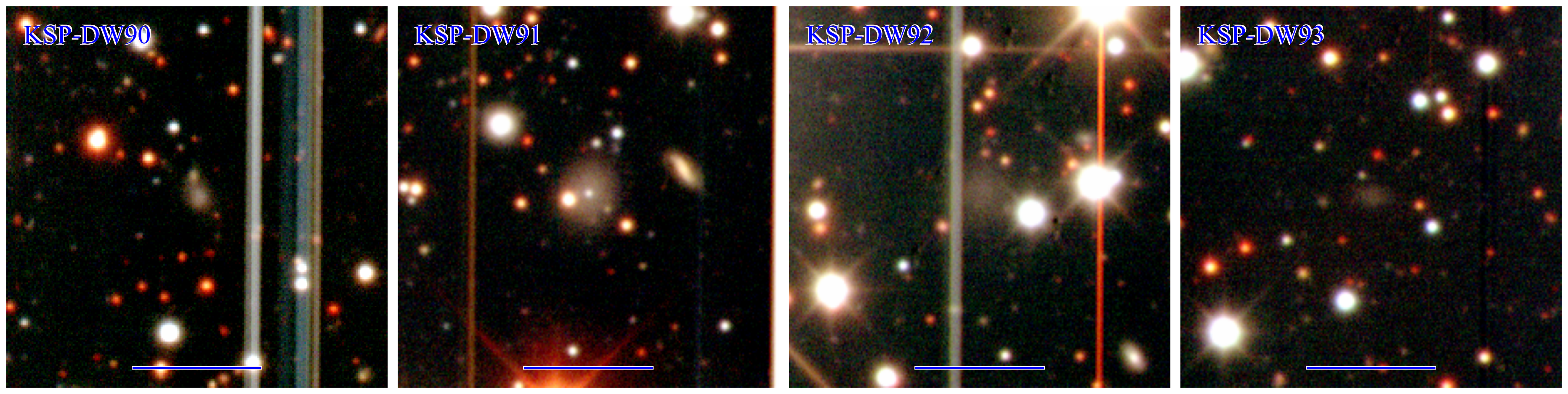}}\\
{\includegraphics[width = \textwidth]{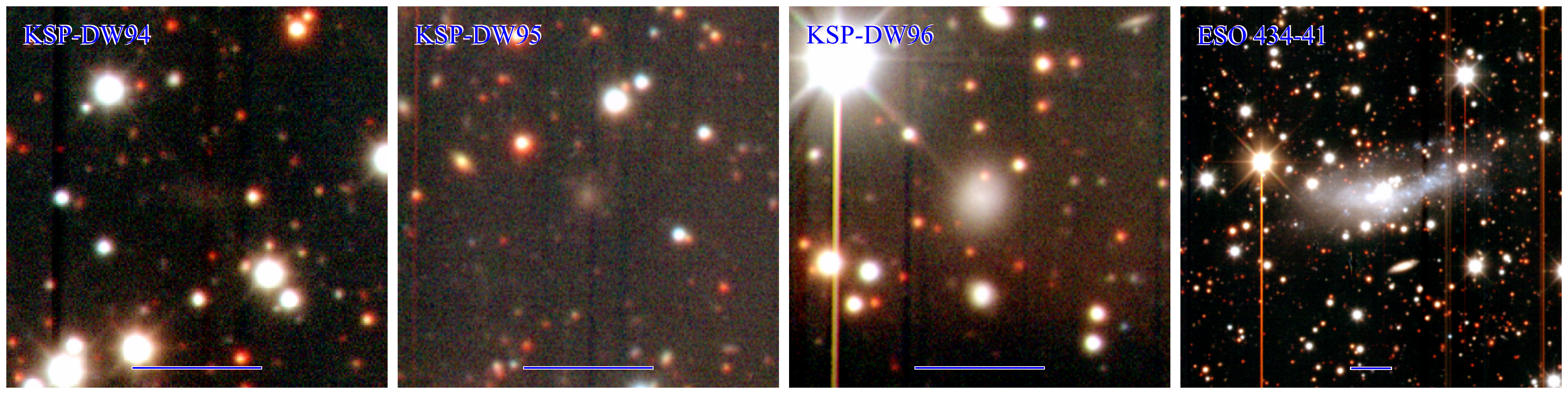}}\\
{\includegraphics[width = \textwidth]{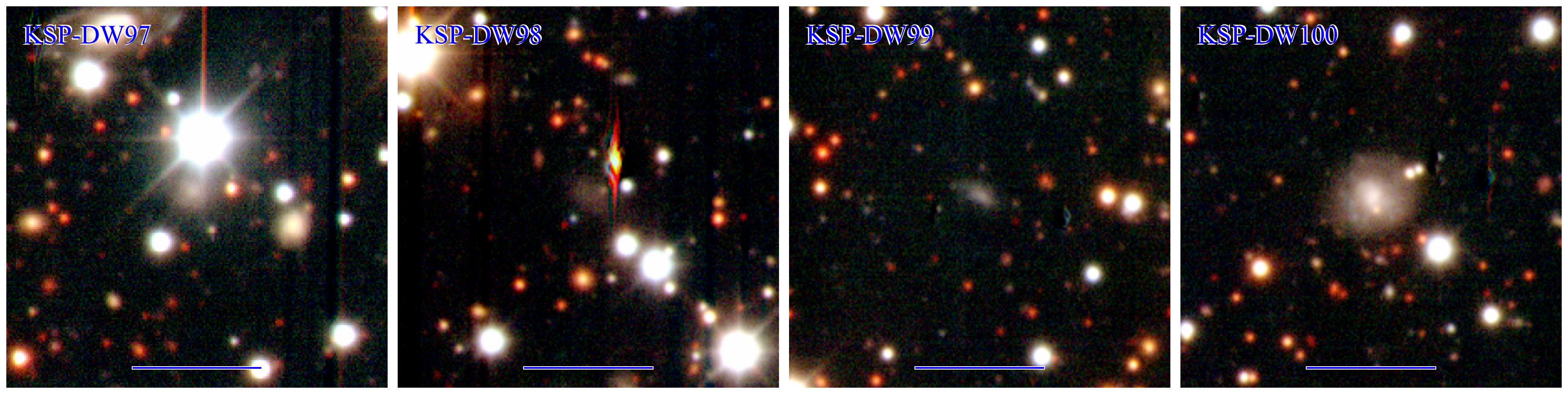}}\\
{\includegraphics[width = \textwidth]{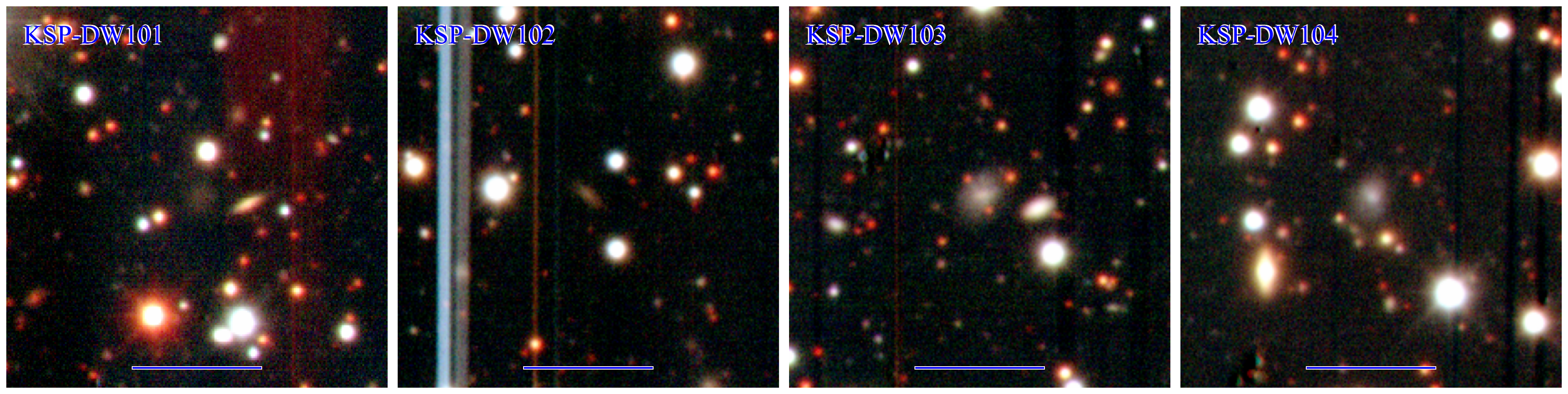}}\\
\centerline{Fig.~\ref{fig:cutoutfull}. --- {\it Continued}}
\end{figure*}

\begin{figure*}
\center
{\includegraphics[width = \textwidth]{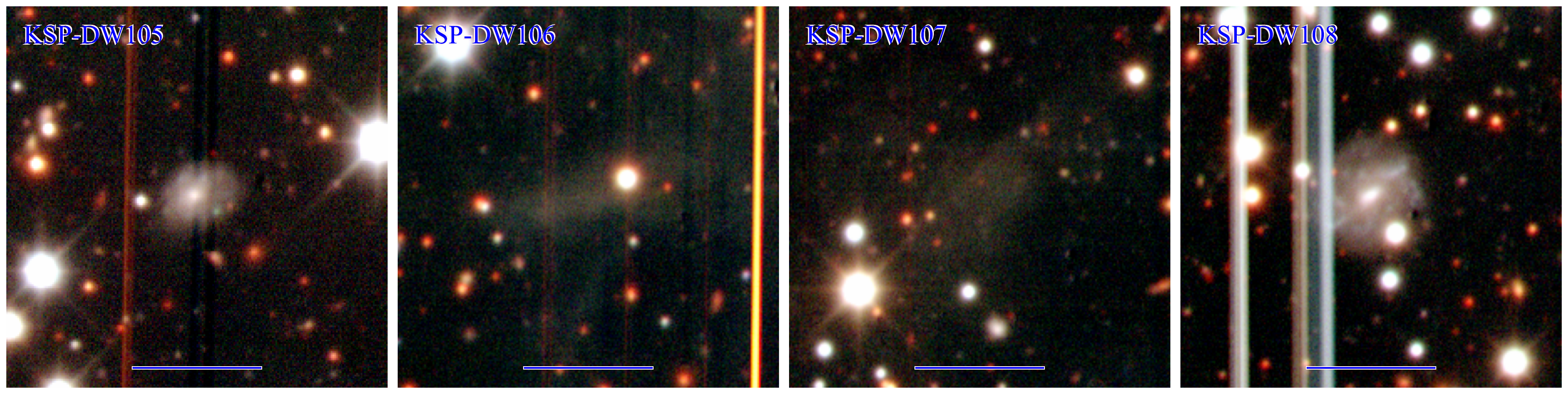}}\\
{\includegraphics[width = \textwidth]{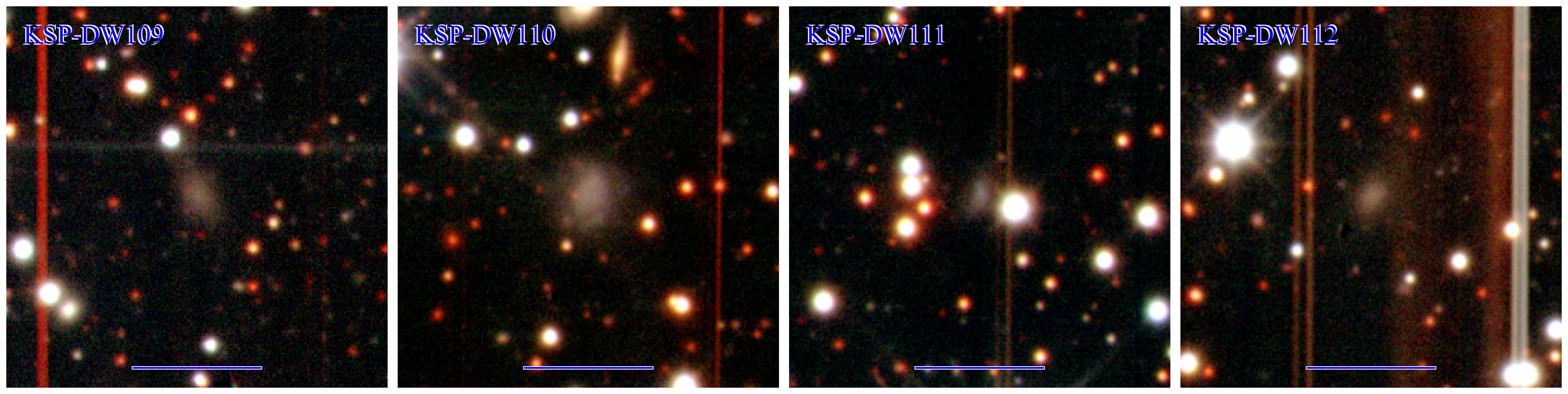}}\\
{\includegraphics[width = \textwidth]{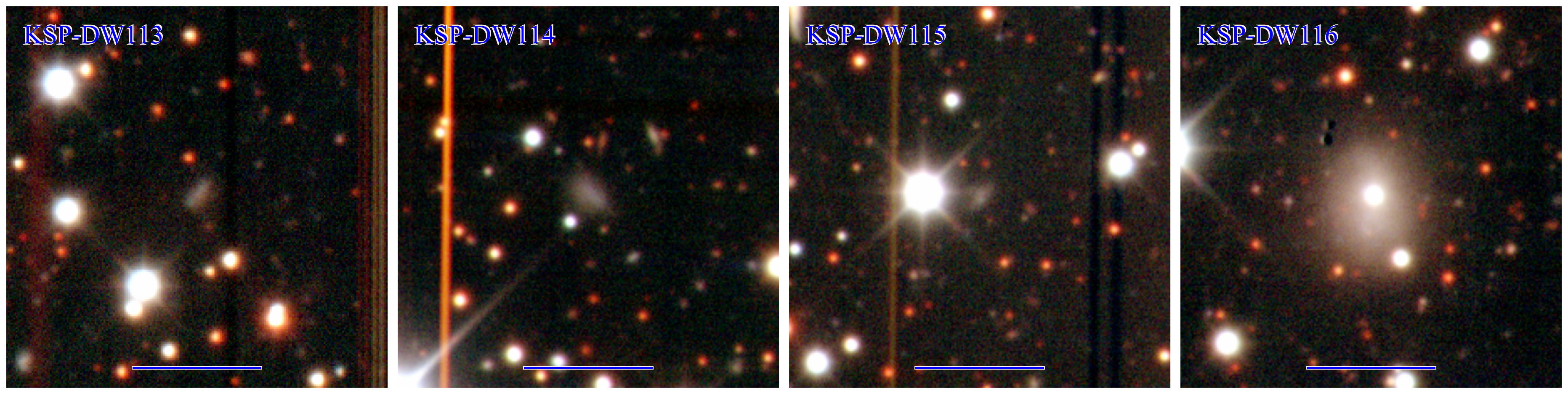}}\\
{\includegraphics[width = \textwidth]{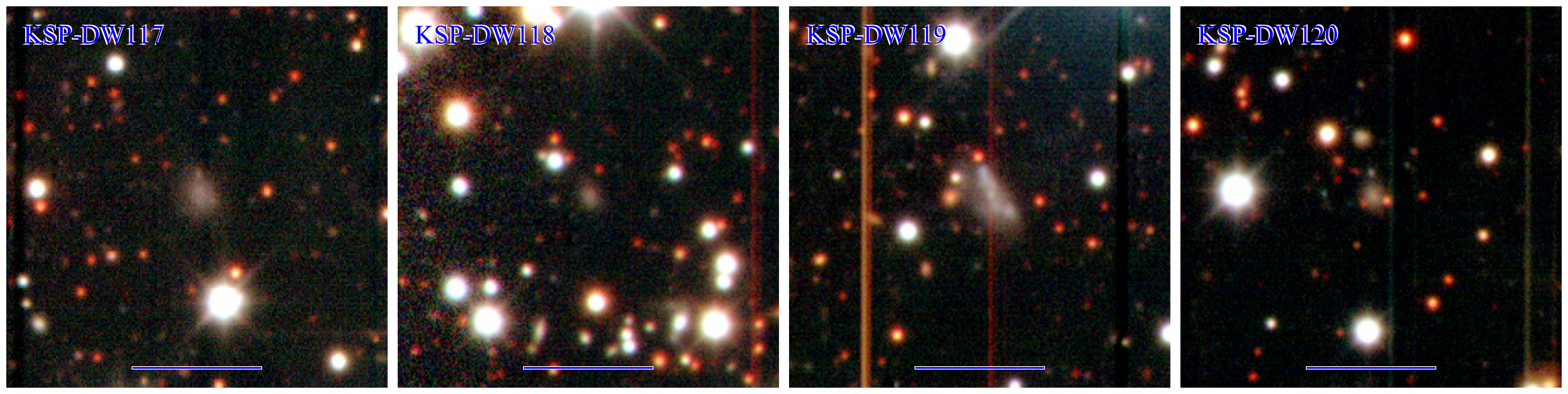}}\\
\centerline{Fig.~\ref{fig:cutoutfull}. --- {\it Continued}}
\end{figure*}

\begin{figure*}
\center
{\includegraphics[width = \textwidth]{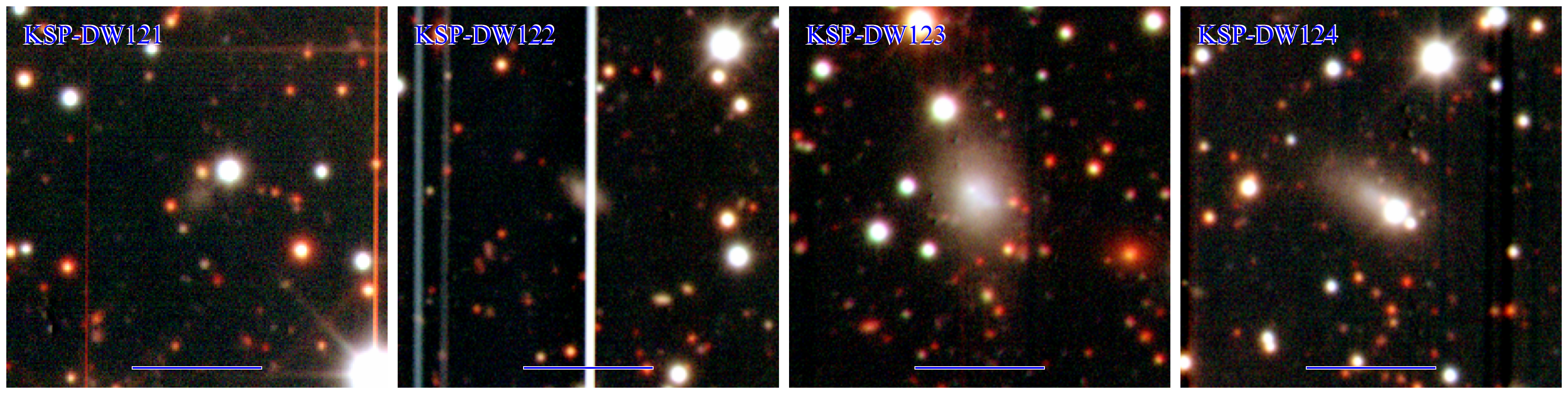}}\\
{\includegraphics[width = \textwidth]{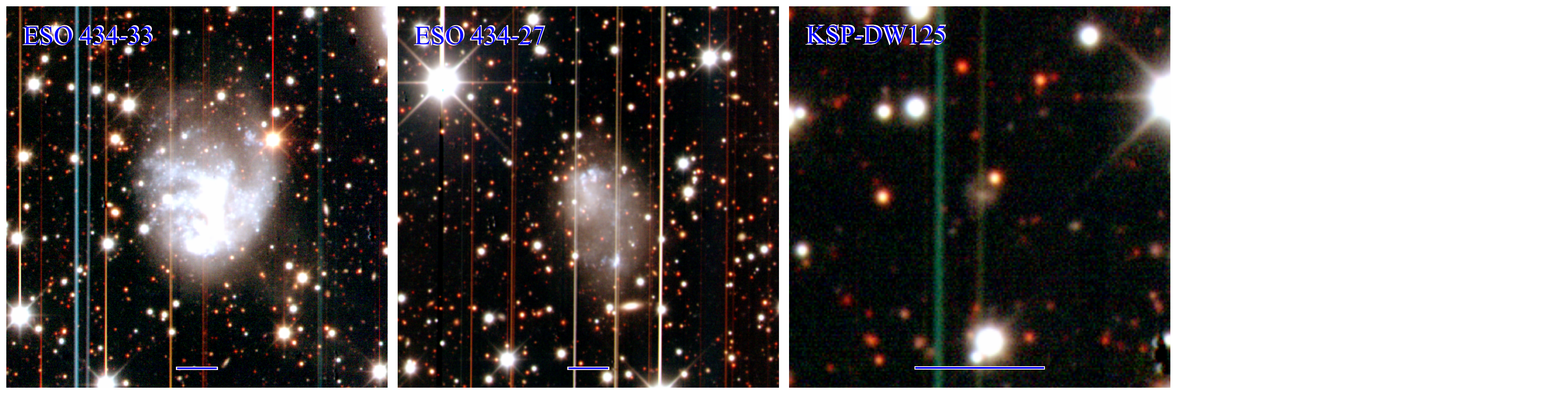}}\\
\centerline{Fig.~\ref{fig:cutoutfull}. --- {\it Continued}}
\end{figure*}

\begin{figure*}
\center
{\includegraphics[width = \textwidth]{Figures/SBP1.pdf}}\\
{\includegraphics[width = \textwidth]{Figures/SBP2.pdf}}\\
{\includegraphics[width = \textwidth]{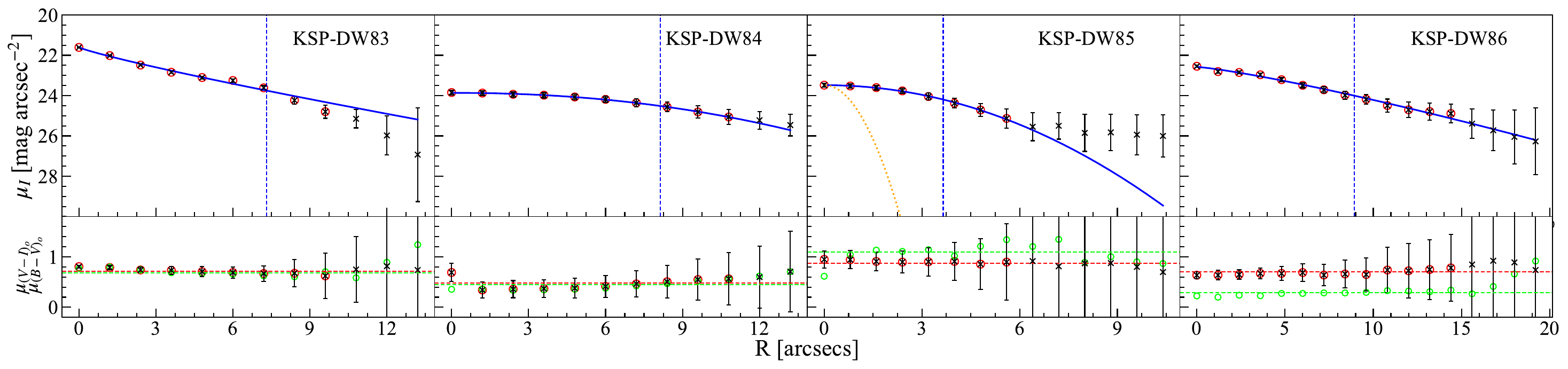}}\\
{\includegraphics[width = \textwidth]{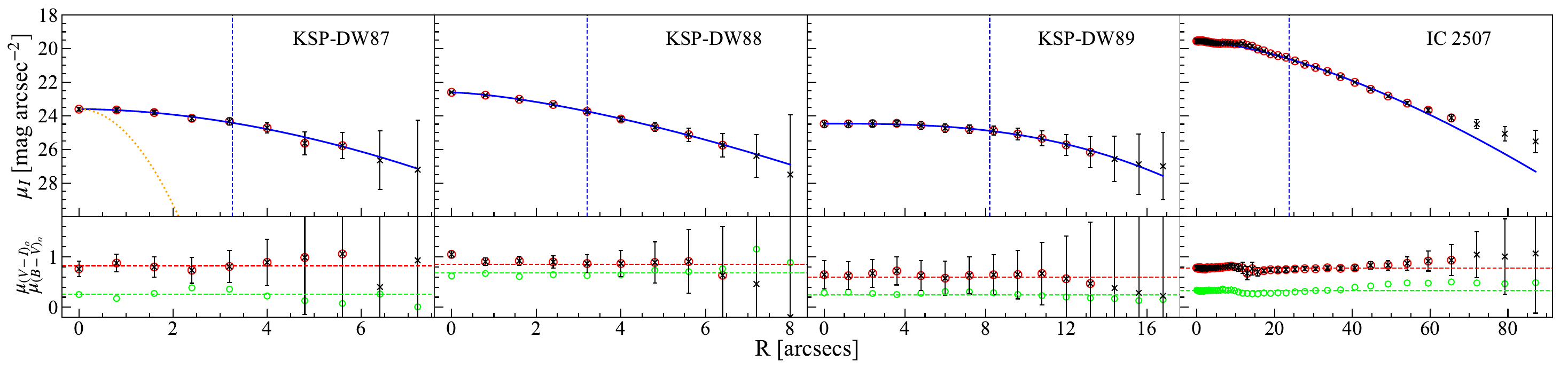}}\\
\caption{{\it Upper Panels}: 
$I$-band surface brightness profiles of the DGCs.
The black crosses with error bars, often encircled by a red circle, 
represent $\mu_I$ and its uncertainty. 
Only those data points with a red circle are used in S\'ersic fitting.
The blue solid curves show the resulting best-fit S\'ersic profiles. 
The vertical blue dashed lines mark the estimated effective radii. 
{\it Lower Panels}: \bvo\ and \vio\ color profiles are shown in 
green circles and black crosses, respectively. 
The green and red horizontal dashed lines represent the mean
\bvo\ and \vio\ colors obtained from difference in total apparent magnitudes
(See \S~\ref{subsec:SBPNCAT}), respectively.
{\tjf The dotted orange curves in the surface brightness profiles of 
KSP-DW78, DW85, DW87, DW102, DW104, DW108, DW115, and DW118 are
the PSF profiles of the images to which the corresponding DGCs belong, 
each representing one image (from F1-Q0 to F2-Q3, see Table~\ref{tab:zero}).} 
\label{fig:SBPFULL}}
\end{figure*}

\begin{figure*}
\center
{\includegraphics[width = \textwidth]{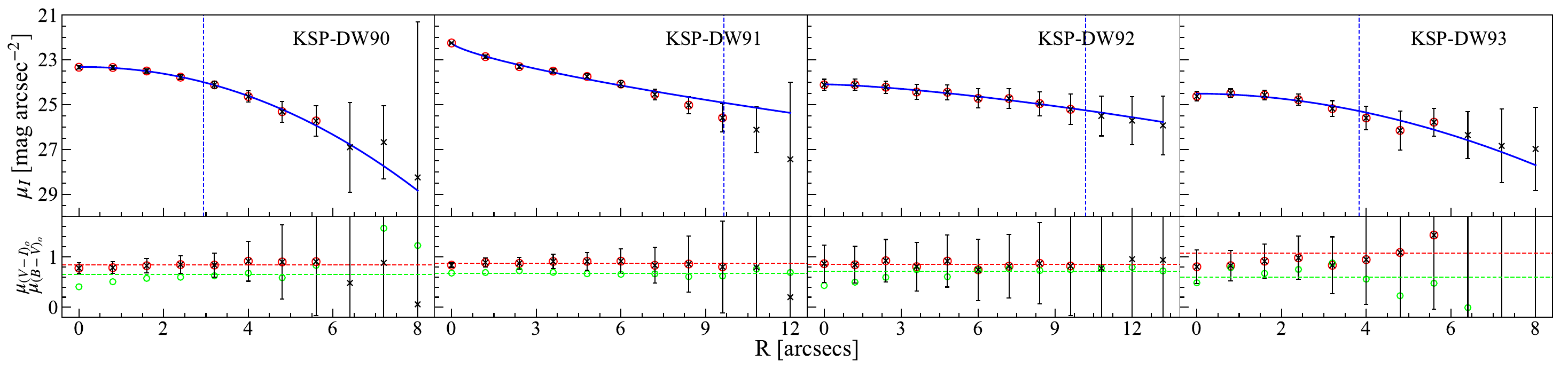}}\\
{\includegraphics[width = \textwidth]{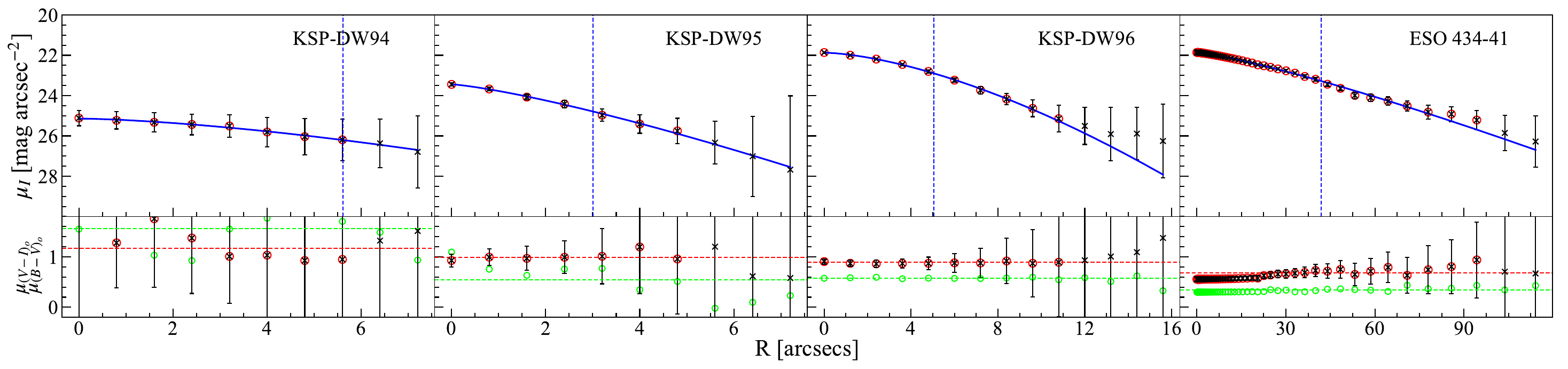}}\\
{\includegraphics[width = \textwidth]{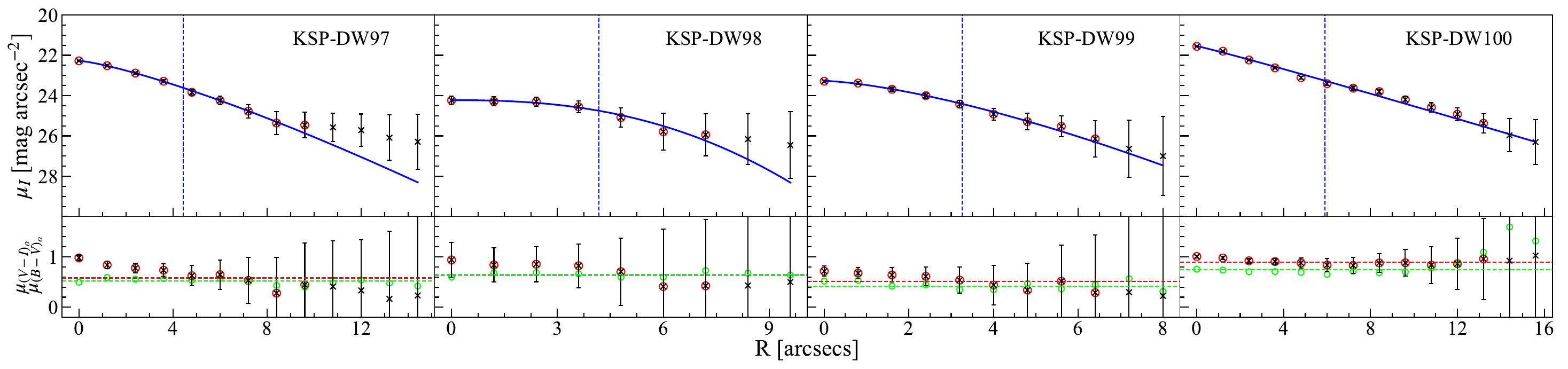}}\\
{\includegraphics[width = \textwidth]{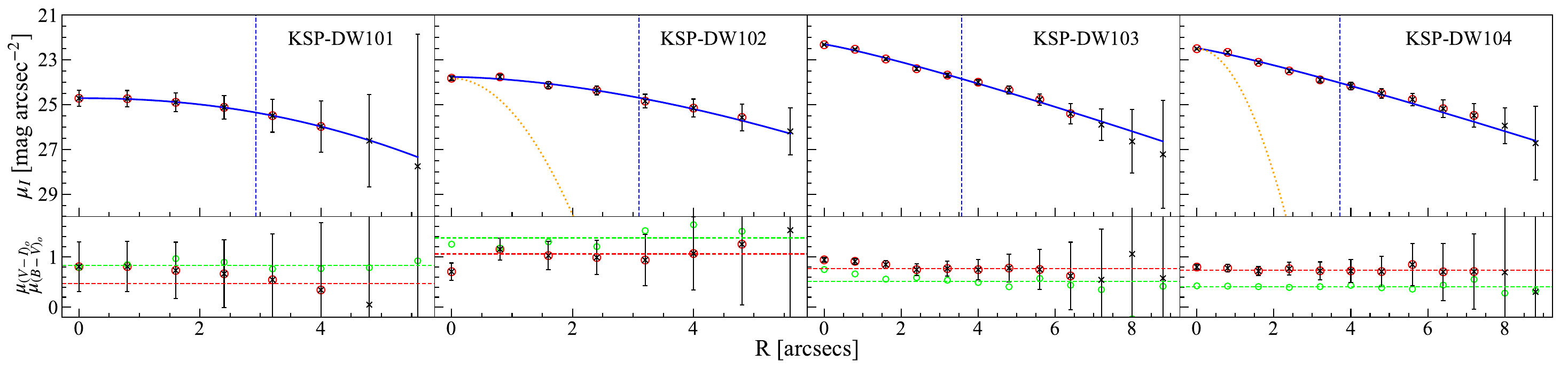}}\\
\centerline{Fig.~\ref{fig:SBPFULL}. --- {\it Continued}}
\end{figure*}

\begin{figure*}
\center
{\includegraphics[width = \textwidth]{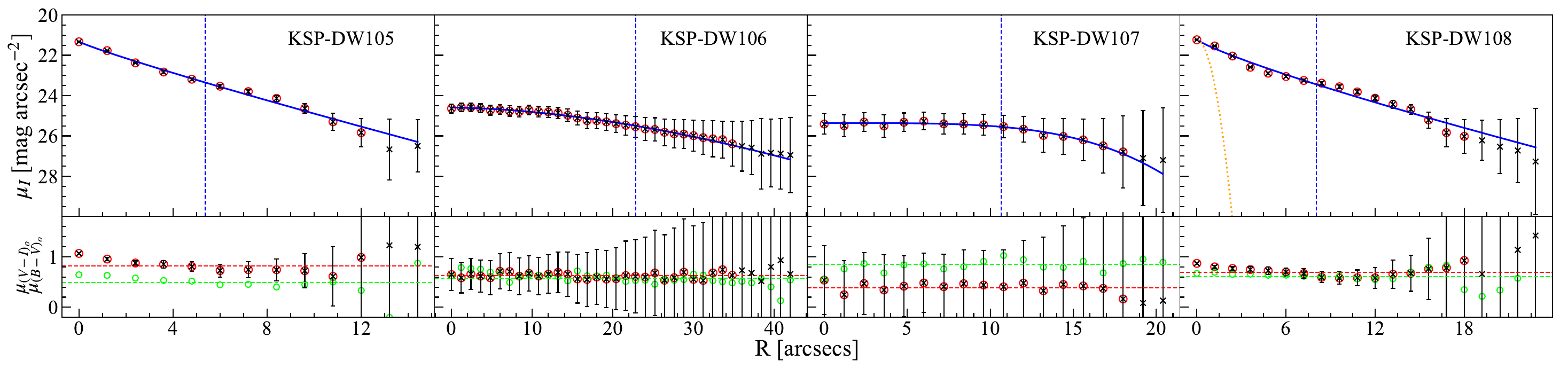}}\\
{\includegraphics[width = \textwidth]{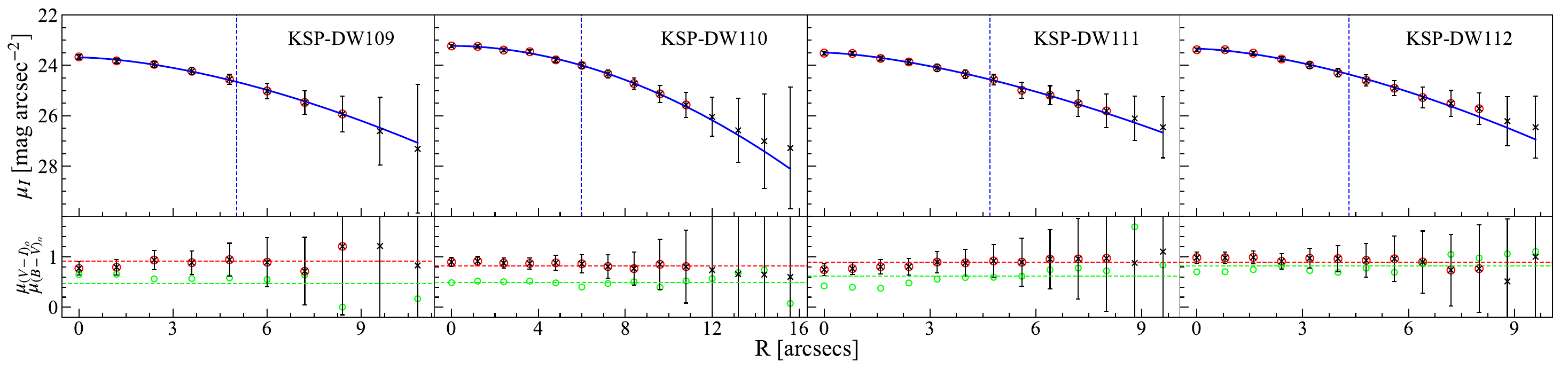}}\\
{\includegraphics[width = \textwidth]{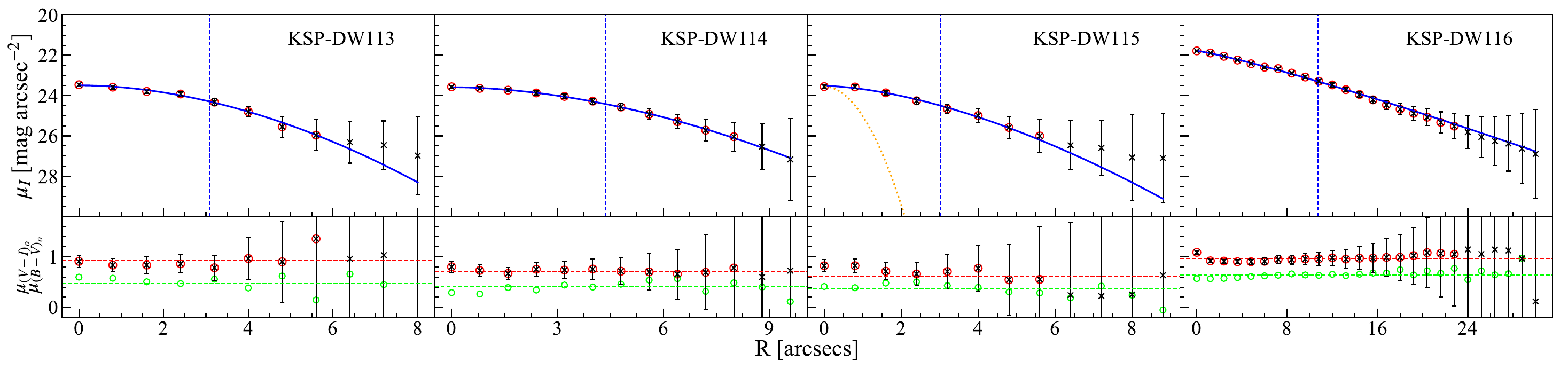}}\\
{\includegraphics[width = \textwidth]{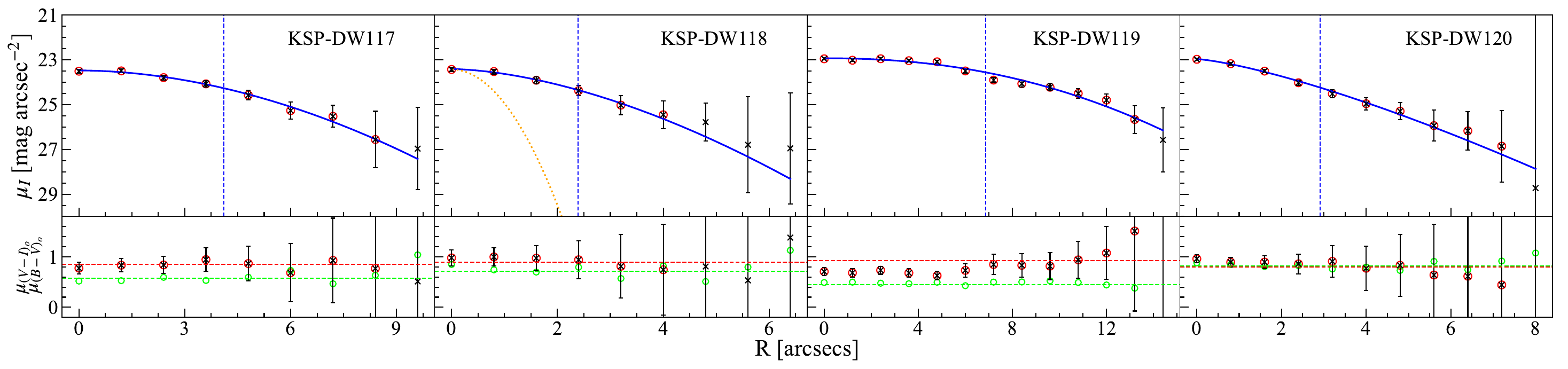}}\\
\centerline{Fig.~\ref{fig:SBPFULL}. --- {\it Continued}}
\end{figure*}

\begin{figure*}
\center
{\includegraphics[width = \textwidth]{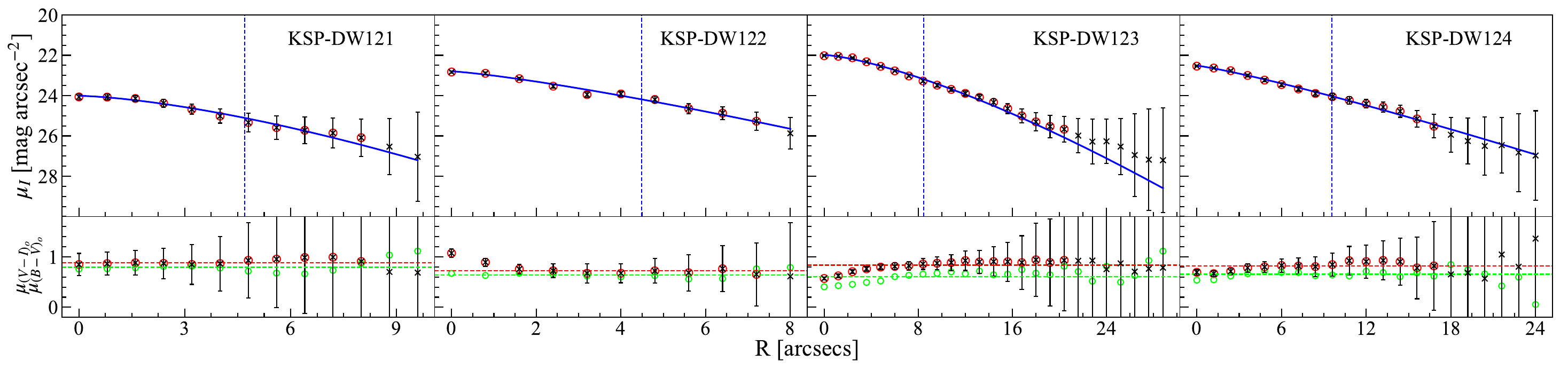}}\\
{\includegraphics[width = \textwidth]{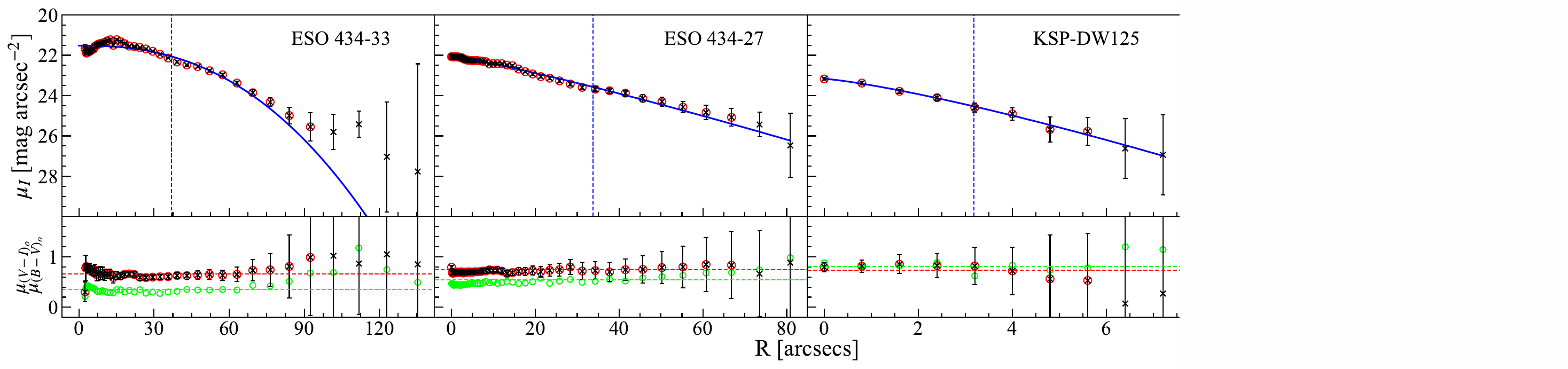}}\\
\centerline{Fig.~\ref{fig:SBPFULL}. --- {\it Continued}}
\end{figure*}

\begin{table*}
\footnotesize
\centering
\begin{threeparttable}[b]
\caption{KMTNet Dwarf Galaxy Discoveries Summary Table}
\label{tab:comp}
\begin{tabular}{ccccccccc}
\hline
Host Name & Host Type & Distance & M$_{B}$ & \# of DGCs{\rm $^a$} & \meanbvo & C{\rm $^b$} & LF Slope\\
          &           & (Mpc)    & (mag)   &                      &   (mag)  &             &         \\
\hline
NGC~2784 & SA(s)    & 9.8  & --19.0 & 38 & 0.66 & --0.12 & --1.22 \\
NGC~3585 & E7/S0    & 20.4 & --21.0 & 50 & 0.69 & --0.06 & --1.33 \\
NGC~2997 & SAB(rs)c & 12.2 & --20.5 & 55 & 0.56 & 0.40   & --1.43 \\
\hline
\end{tabular}
\begin{tablenotes}
\item \rm $^a$The \# of DGCs includes previously discovered dwarf galaxies.
\item \rm $^b$C is the linear correlation coefficient, see \S\ref{subsec:CSP}.
\item Host type and M$_{B}$ information are obtained from Simbad, 
while the rest are from \citealp[]{2017Park} or \citealp[]{2019Park}.
\end{tablenotes}
\end{threeparttable}
\end{table*}


\bsp	
\label{lastpage}
\end{document}